\documentclass[reqno,a4paper]{article}

\pdfoutput=1
\usepackage[top=1in, bottom=1.25in, left=1in, right=1in]{geometry}

\usepackage[ngerman,british]{babel}

\usepackage[utf8]{inputenc}

\usepackage{soul}

\usepackage[T1]{fontenc}
\usepackage{lmodern}
\usepackage{amsmath,amssymb,amsthm,amsfonts}
\usepackage{lineno}
\usepackage{mathtools}
\mathtoolsset{showonlyrefs}
\usepackage{nicefrac}       
\usepackage{braket}
\usepackage{xspace}
\usepackage{float}
\restylefloat{table}
\usepackage{booktabs}
\usepackage{bibentry}
\usepackage{authblk}
\usepackage{subcaption}
\usepackage{graphicx}
\usepackage{color}
\usepackage{epstopdf}
\DeclareGraphicsExtensions{.pdf,.eps,.png}
\usepackage[export]{adjustbox}
\usepackage{stackengine}
\usepackage{tikz}
\usetikzlibrary{shapes.misc}

\usepackage{paralist}
\usepackage{enumitem}

\usepackage{bbm}

\usepackage{calc}

\usepackage{accents}

\usepackage{todonotes}

\usepackage{geometry}

\usepackage[bookmarksnumbered,bookmarksopen,unicode,colorlinks,allcolors=purple]{hyperref}
\usepackage{url}
\usepackage{algorithm}
\usepackage{algpseudocode}
\usepackage[
    doi=true,
    url=true,
    isbn=false,
    maxbibnames=999,
    eprint=true,
    backend = biber,
    sorting=none,
    sortcites=true
    ]{biblatex}
\AtEveryBibitem{%
  \clearlist{language}
}

\title{Quantum Portfolio Optimization: An Extensive Benchmark}

\author[1,*]{Eric Stopfer}
\author[1]{Friedrich Wagner}
\affil[1]{Fraunhofer Institute for Integrated Circuits, Nürnberg}
\affil[*]{\texttt{\small eric.stopfer@iis.fraunhofer.de}}

\date{October 2026}

\begin{document}
	
	\maketitle
	
	\begin{abstract}
 Recently, several researchers proposed portfolio optimization as a potential use case for quantum optimization.
 However, the literature is lacking an extensive benchmark quantifying the potential of quantum computers for portfolio optimization.
 In this work, we contribute to closing this gap.
 We provide a computational study, comparing quantum approaches against state-of-the-art classical methods on a meaningful, real-world instance set.
 In particular, we compare quantum annealing and the quantum approximate optimization algorithm against classical mixed-integer programming, simulated annealing, steepest descent local search, tabu search and a problem-tailored heuristic.
 We consider a variance-minimizing variant of portfolio optimization which we show to be more difficult to solve for classical optimizers than return-maximizing or multi-objective formulations.
 Our benchmark data set comprises 270 instances with up to 1,000 assets from actual stock data.
 Due to hardware limitations, quantum methods could only be tested for instances with at most 30 assets.
 The results show that all instances can be solved to proven optimality by mixed-integer programming in the order of seconds.
 Moreover, the problem-tailored heuristic consistently outperforms quantum approaches in terms of solution quality for fixed runtime.
 Thus, we conclude that there is only very limited room for a potential quantum advantage for the considered variant of portfolio optimization.
\end{abstract}

\section{Introduction}
Combinatorial optimization problems play a central role in various fields such as logistics and finance~\cite{korte_combinatorial_2018}.
In practice, these problems are often tackled by mathematical optimization solvers, which can solve large-scale mixed-integer and quadratic optimization problems, including many NP-hard instances,
to proven optimality~\cite{koch_progress_2022, junger_50_2010}.
Nevertheless, some practically relevant problems remain intractable for state-of-the-art classical methods~\cite{gleixner_miplib_2021}.
This fact has motivated researchers to develop quantum algorithms for such problems~\cite{abbas_challenges_2024}.
However, the limitations of existing quantum computers prohibit the implementation of exact quantum algorithms for optimization problems~\cite{ammann_realistic_2023}.
As a result, researchers focus on heuristic quantum algorithms, which require significantly less resources than exact approaches~\cite{bochkarev_quantum_2026}.
Well-known examples of such quantum heuristics are quantum annealing~\cite{albash_demonstration_2018, mcgeoch_milestones_2023} and the quantum approximate optimization algorithm (QAOA)~\cite{farhi_quantum_2014, harrigan_quantum_2021}.
Recently, several works proposed portfolio optimization
as a suitable problem class for achieving quantum advantage~\cite{koch_quantum_2026,sakuler_real_2023,liu_hybrid_2022,acharya_decomposition_2025, phillipson_portfolio_2021,palmer_quantum_2021, venturelli_reverse_2019,brandhofer_benchmarking_2022, hodson_portfolio_2019,tang_comparative_2024,buonaiuto_best_2023}.
In this work, we provide experimental evidence
suggesting that even large instances of portfolio optimization can be solved to proven optimality by modern classical solvers in the order of seconds.
Moreover, our problem-tailored classical heuristic outperforms quantum approaches in terms of solution quality for fixed runtime.
Our results thus set the bar high for a practical quantum advantage for the considered variant of portfolio optimization.

\paragraph{Related work.}
Various variants of the portfolio optimization problem exist in literature.
The authors of~\cite{brandhofer_benchmarking_2022, acharya_decomposition_2025, venturelli_reverse_2019, phillipson_portfolio_2021} consider a variant that can be described as \emph{choose-asset-or-not}.
Therein, as the name suggests, the task is to decide for each asset if we add it to the portfolio or not, equally weighted.
In this work, we consider an extended variant
where the goal is to find optimal \emph{weights} of assets in the portfolio, that is,
we assign a fraction of the invested money to each asset.
This variant can be further subdivided with respect to the considered objective.
Some researchers focus on maximizing return with limited return variance~\cite{sakuler_real_2023,palmer_quantum_2021}.
Others aim to minimize the variance with a fixed return~\cite{venturelli_reverse_2019, cesarone_efficient_nodate}.
Several authors combine both objectives by maximizing a weighted sum of return and variance~\cite{brandhofer_benchmarking_2022,acharya_decomposition_2025,rubio-garcia_portfolio_2022}.
There also exist approaches in literature
which include further practical aspects of stock trading into the model,
for example, short-selling-options, risk-free investments and costs that are associated with asset acquisitions~\cite{koch_quantum_2026, rubio-garcia_portfolio_2022, hodson_portfolio_2019}.
 The authors of~\cite{koch_quantum_2026} consider a data test set consisting of $10$, $50$, $200$ and $400$ assets of the S\&P500 stock market which is similar in structure but less granular than the data used in this work.

Classical solution approaches to the portfolio optimization problem
include both heuristics~\cite{chang_heuristics_2000, buhler_efficient_2023, rubio-garcia_portfolio_2022, alessandroni_alleviating_2025} and exact methods~\cite{moeini_continuous_2014, cesarone_portfolio_2015, moka_scalable_2025}.
Notably, for fixed runtime, the heuristic developed in~\cite{moka_scalable_2025} finds solutions of better quality for instances of the choose-asset-or-not problem than the commercial classical solver CPLEX~\cite{noauthor_ibm_2022}.

Recently, researchers proposed quantum approaches to the portfolio problem, including QAOA~\cite{brandhofer_benchmarking_2022, hodson_portfolio_2019}, Variational Quantum Eigensolver (VQE)~\cite{buonaiuto_best_2023}
and quantum annealing~\cite{sakuler_real_2023, palmer_quantum_2021, venturelli_reverse_2019, phillipson_portfolio_2021,tang_comparative_2024}.
The authors of~\cite{sakuler_real_2023},~\cite{phillipson_portfolio_2021} and~\cite{palmer_quantum_2021} conclude that quantum annealing may outperform classical methods for portfolio optimization.
However, their studies employ a closed-source hybrid quantum-classical solver,
which makes it impossible to divide classical from quantum contributions to the solution.
On the contrary, in this work, we stick to methods with known classical and quantum components.
Regarding QAOA, the authors of~\cite{hodson_portfolio_2019} report near-optimal results for small problem instances of 8 assets with an idealized simulator of a gate-based quantum computer.
Ref.~\cite{brandhofer_benchmarking_2022} provides guidelines on the selection of penalty and circuit parameters as well as on the classical optimizer.
In~\cite{venturelli_reverse_2019, tang_comparative_2024}, the authors use reverse quantum annealing to improve upon conventional quantum annealing.
Ref.~\cite{buonaiuto_best_2023} studies the influence of penalty coefficients in VQE as well as the effects of different quantum devices.
Other approaches cover problem sparsification~\cite{buhler_efficient_2023}, problem decomposition~\cite{acharya_decomposition_2025}, warm-starts~\cite{schlütter2025hotstartingquantumportfoliooptimization} and quantum circuit cutting techniques~\cite{soloviev_scaling_2025}.
For a comprehensive review on solution approaches to the portfolio optimization, we refer the interested reader to~\cite{loke_portfolio_2023}.

\paragraph{Our contribution.}
In this work, we provide a computational study, including $270$ problem instances with up to $1,000$ assets.
The instances are generated from real-world stock data 
providing instances with realistic return and covariance. However, we note that we do not make any tests based on actual investment performance.
Moreover, we find optimal portfolio weights instead of only deciding whether to select an asset or not.
Thus, our approach resembles practical scenarios.
We compare QAOA, quantum annealing, two exact classical solvers, three meta-heuristics and one problem-specific heuristic.
Our contribution is twofold.
First, the large instance set allows us to draw conclusions that are
consistent across different instances and instance sizes.
Second, applying standard implementations of state-of-the-art quantum and classical solution methods, we provide a comprehensive and fair benchmark.
In particular, our methods have well-defined quantum and classical parts such that we can precisely divide quantum from classical contributions.

The remainder of this work is organized as follows.
In section~\ref{sec:original_problem}, we define three variants of portfolio optimization which we consider in this work.
Section~\ref{sec:classical_sol_time_minvola_maxret_multiobj} studies the classical hardness of these variants to identify the one with the largest potential for quantum advantage.
In section~\ref{sec:qubo_transformation}, we develop a quadratic unconstrained binary optimization (QUBO) model for the most difficult portfolio optimization variant.
Section~\ref{sec:methods} introduces the solution methods we compare in our benchmark.
In Section~\ref{sec:benchmark_study_results}, we report on the results of our benchmark study.
Finally, in section~\ref{sec:conclusion}, we summarize our findings, draw a conclusion and state open questions.

\section{The Portfolio Optimization Problem}\label{sec:original_problem}
In this section, we introduce three different variants
of the portfolio optimization problem, which we consider in our computational study.
As originally formulated by Markowitz in~\cite{markowitz_portfolio_1952},
the portfolio optimization problem asks for a selection of an asset portfolio which maximizes the expected portfolio return while
minimizing the variance of the return..

More formally, given a set of $n$ assets, the task is to find optimal \emph{asset weights} $\omega_i \in [0,1]$, $i\in \{1,\dots,n\}$.
The asset weight $\omega_i$ defines the share of the invested money allocated to the $i$-th asset.
Accordingly, it holds
\begin{equation}
 \sum_{i=1}^n \omega_i = 1\ .
\end{equation}
We impose upper bounds $u_i \in (0,1]$ on the asset weights,
\begin{equation}
 \omega_{i} \leq u_i\ \quad \forall i \in {1,\dots, n}.
\end{equation}
Each asset $i$ has an \emph{expected return} $r_i \in \mathbb{R}$.
Moreover, for each pair of assets $(i,j)$, we are given a \emph{return covariance} $\sigma_{ij} \in \mathbb{R}$.

The original problem formulation by Markowitz~\cite{markowitz_portfolio_1952} includes the objectives of maximizing the expected portfolio return
\begin{equation}
 \mu(\omega) \coloneqq \sum_{i=1}^n \omega_i r_i
\end{equation}
and minimizing the portfolio variance
\begin{equation}
 \sigma^2(\omega) \coloneqq \sum_{i=1}^n \sum_{j=1}^n \omega_i \omega_j \sigma_{ij}\ .
\end{equation}
In general, the two objectives of maximizing return and minimizing variance cannot be fulfilled at the same time.
Thus, several problem variants exist, which we summarize in the following.
In the first variant, we aim at maximizing the return with an upper bound $\nu > 0$ on the variance.
This can be formulated by the quadratic program
\begin{align}
 \hypertarget{eq:max_ret}{\textbf{MaxRet}: \quad}\max_{\omega} \quad & \mu(\omega) \\
 \text{s.t.} \quad & \sigma^2(\omega) \leq \nu \\
 & \sum_{i = 1}^{n} \omega_{i} = 1 \\
 & 0 \leq \omega_{i} \leq u_i \quad \forall i \in \{1,\dots,n\}\ .
\end{align}
The second variant minimizes the variance while ensuring a minimum return $\epsilon \in \mathbb{R}$.
This is formalized by the quadratic program
\begin{align}
 \hypertarget{eq:min_vola}{\textbf{MinVola}: \quad} \min_{\omega} \quad & \sigma^2(\omega) \\
 \text{s.t.} \quad & \mu(\omega) \geq \epsilon \label{eq:minreturn_constraint} \\
 & \sum_{i = 1}^{n} \omega_{i} = 1 \label{eq:normalization_constraint}\\
 & 0 \leq \omega_{i} \leq u_i \quad \forall i \in \{1,\dots,n\}\ .
\end{align}
The name of the QP reflects its objective of minimizing portfolio volatility.
Finally, the third variant combines both objectives by a fixed factor $\lambda > 0$, which depends on the risk aversion of the investor.
This gives rise to the quadratic program
\begin{align}
 \hypertarget{eq:multi_ob}{\textbf{MultiObj}: \quad} \min_{\omega} \quad & \sigma^2(\omega) - \lambda \mu(\omega) \\
 \text{s.t.} \quad & \sum_{i = 1}^{n} \omega_{i} = 1 \\
 & 0 \leq \omega_{i} \leq u_i\quad \forall i \in \{1,\dots,n\}\ .
\end{align}
The mathematical programs \hyperlink{eq:max_ret}{\text{MaxRet}}, \hyperlink{eq:min_vola}{\text{MinVola}} and \hyperlink{eq:multi_ob}{\text{MultiObj}} belong to the class of convex quadratically constrained quadratic programs, which can be solved in polynomial time by interior point methods~\cite{boyd_convex_2004, wright_interior-point_2004}.
The convexity is due to the fact that $(\sigma_{ij}) \in \mathbb{R}^{n\times n}$ is a sample covariance and thus positive semidefinite by definition~\eqref{eq:def_sigma} in Appendix~\ref{sec:generating_stock_data}.
Although being efficiently solvable in theory, portfolio optimization is often tackled via heuristics in practice.

\paragraph{ Test set generation.} \label{sec:testset_description}
In order to generate a realistic set of instances,
we consider the daily prices of $1,978$ assets in the Nasdaq stock exchange~\cite{noauthor_nasdaq_nodate} during the years 2020 to 2023,
available in~\cite{noauthor_yahoo_nodate} and accessible with~\cite{noauthor_yfinance_2025}.
All those assets have no missing price data during this timeframe.
From the daily prices, we calculate the expected returns and the return covariances.
For details on the calculation of returns and covariances from price data, we refer to Appendix~\ref{sec:generating_stock_data}.
To generate an instance with a given number $n$ of assets, we randomly draw $n$ assets from the data base.

We generate $10$ instances for each problem variant \hyperlink{eq:max_ret}{\text{MaxRet}}, \hyperlink{eq:min_vola}{\text{MinVola}} and \hyperlink{eq:multi_ob}{\text{MultiObj}} for each size $n\in \{3,\allowbreak 5,\allowbreak 7,\allowbreak 8,\allowbreak 10,\allowbreak 15,\allowbreak 20,\allowbreak 25,\allowbreak 30,\allowbreak 40,\allowbreak 50,\allowbreak 60,\allowbreak 70,\allowbreak 80,\allowbreak 90,\allowbreak 100,\allowbreak 150,\allowbreak 200,\allowbreak 250,\allowbreak 300,\allowbreak 400,\allowbreak 500,\allowbreak 600,\allowbreak 700,\allowbreak 800,\allowbreak 900,\allowbreak 1000\}$.
We still need to define values for the maximum portfolio variance $\nu$ in \hyperlink{eq:max_ret}{\text{MaxRet}}, the minimum portfolio return $\epsilon$ in \hyperlink{eq:min_vola}{\text{MinVola}} and the risk aversion $\lambda$ in \hyperlink{eq:multi_ob}{\text{MultiObj}}.
As we strive towards an above-average performing portfolio, in Model~\hyperlink{eq:max_ret}{\text{MaxRet}}, we choose the maximum portfolio variance $\nu$ as the $30$ $\%$ quantile of variances of randomly generated portfolios.
Similarly, in Model~\hyperlink{eq:min_vola}{\text{MinVola}}, we choose the minimum portfolio return $\epsilon$ as the $70\ \%$ quantile
of all asset returns.
Moreover, in order to account for the different magnitudes of $\mu$ and $\sigma^2$, we set the risk aversion factor $\lambda$ of Model~\hyperlink{eq:multi_ob}{\text{MultiObj}} to the average of fraction $\frac{\sigma^2}{\mu}$ of randomly generated portfolios.
We remark that, in general, the selection of the parameters $\epsilon,\nu, \lambda$ is subject to the investor's perspective.
Finally, we set the asset limits to $u_i=\max\left(\frac{1}{10}, \frac{3}{n}\right)$ for $i\in\{1,...,n\}$.
The asset limits thus form a monotonically decreasing series, which for 5 assets has a value of 60$\%$, for 10 assets 30$\%$, for 20 assets 15$\%$ and stays at 10$\%$ from $n=30$ onward.
By this formula, we try to emulate real-world fund compositions, which can be monitored in~\cite{noauthor_yahoo_nodate}.

In principle, some generated test instances of~\hyperlink{eq:min_vola}{\text{MinVola}} and~\hyperlink{eq:max_ret}{\text{MaxRet}} could be infeasible.
However, our computational experiments revealed that all considered test instances are indeed feasible.

\section{Classical Solution Time for Different Variants}\label{sec:classical_sol_time_minvola_maxret_multiobj}
In this section, we study the classical computational difficulty of the three problem variants~\hyperlink{eq:max_ret}{\text{MaxRet}}, ~\hyperlink{eq:min_vola}{\text{MinVola}} and~\hyperlink{eq:multi_ob}{\text{MultiObj}}.
Our goal is to identify the hardest variant for classical solvers which thus bears the largest potential for a quantum advantage.
To this end, we solve all instances of the test set described in Section~\ref{sec:testset_description} with classical mathematical optimization solvers to proven optimality and compare the overall runtime.
For each instance, we solve the three problem variants with the open-source solver SCIP 9.2~\cite{bolusani_scip_nodate, noauthor_pyscipopt_2025} and the commercial solver Gurobi 11.0.3~\cite{noauthor_leader_nodate}.
Moreover, we impose a time limit of $3,600$~s and relative optimality gaps of $0\ \%$ and $5\ \%$.
Here, the relative optimality gap, short gap, is defined as
\begin{align}\label{eq:gap}
 g\coloneqq\frac{|b-c|}{|c|}
\end{align}
where $b$ is the best bound on the objective and $c$ is the objective value of the best available solution.

\begin{figure}[]
 \centering
 \begin{subfigure}{0.48\textwidth}
 \centering
 \includegraphics[width=0.98\linewidth]{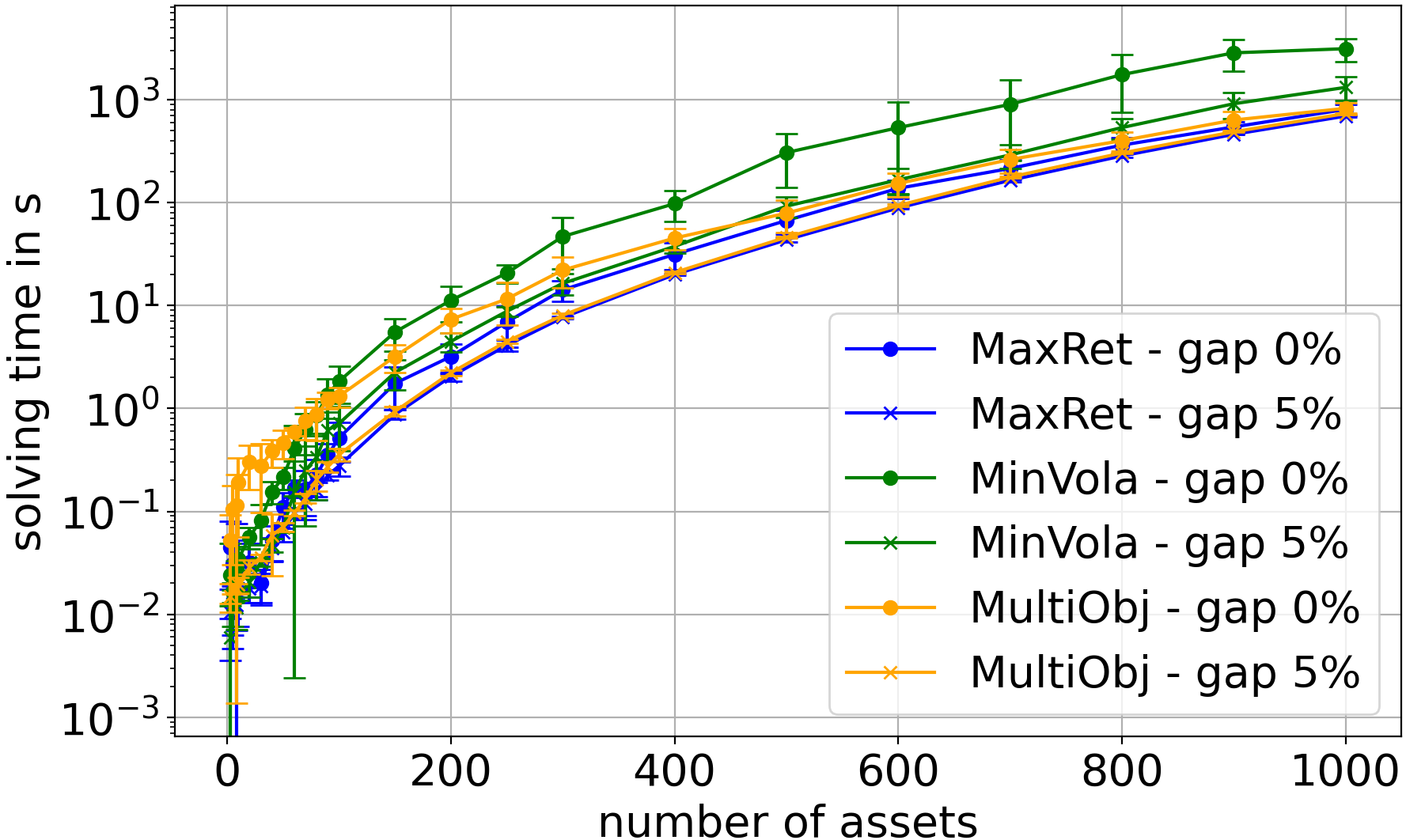}
 \caption{SCIP}
 \label{solving_times_scip}
 \end{subfigure}
 \begin{subfigure}{0.48\textwidth}
 \centering
 \includegraphics[width=0.98\linewidth]{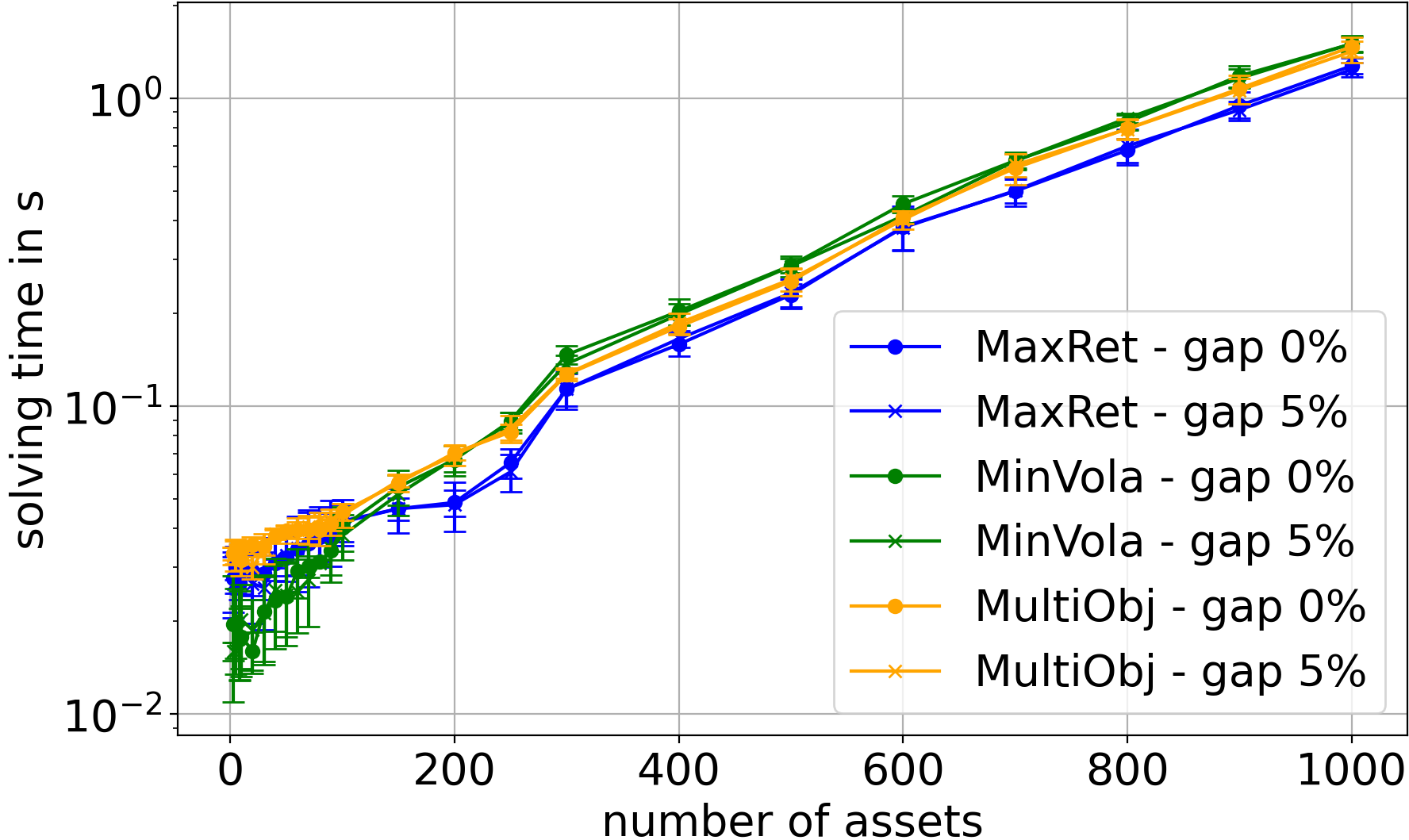}
 \caption{Gurobi}
 \label{solving_times_gurobi}
 \end{subfigure}
 \caption{Average solver runtime for different problem variants. We compare SCIP (a) and Gurobi (b) with relative optimality gaps of $0$ \% and $5$ \%. Data points are averages over $10$ instances and error bars show the empirical standard deviation.}
 \label{fig:solution_times}
\end{figure}

The average solving times are visualized in Figure~\ref{fig:solution_times}.

First, we observe that Gurobi solves most of the problems in less than one second and thus is considerably faster than the open-source solver SCIP, with a speed-up factor of over $1,000$ for large problem instances with $n=1,000$ assets.
The speed-up can partly be explained by the fact that Gurobi detects the convexity of the problem and consequently solves it using the Barrier algorithm, a polynomial-time interior-point method.
In contrast, as observed in the solver logs, SCIP relies on outer approximation with spatial branching, which is a method for general, non-convex QPs and has exponential worst case running time.

Second, the influence of the duality gap on the solving time is larger for SCIP than for Gurobi.

Third, the Model~\hyperlink{eq:min_vola}{\text{MinVola}} takes the longest time to be solved in each of the four solver-gap configurations.
Thus, \hyperlink{eq:min_vola}{\text{MinVola}} is the most promising variant for a potential quantum speedup.
Consequently, in the following we focus on the~\hyperlink{eq:min_vola}{\text{MinVola}} variant that minimizes variance while still maintaining a certain portfolio return level.

\section{QUBO Transformation}\label{sec:qubo_transformation}
Existing implementations of both quantum annealing and QAOA require the input problem to
be modeled as QUBO,
that is, problems of the form
\begin{align}\label{eq:QUBO}
 \min_{x\in\{0,1\}^n} x^tQx
\end{align}
where $n\in \mathbb{N}$ and $Q\in \mathbb{R}^{n\times n}$ is a real-valued matrix.
In this section, we derive a QUBO model of the \hyperlink{eq:min_vola}{\text{MinVola}} variant.
To this end, we define the penalty-based objective function
\begin{equation}\label{eq:qubo_obj}
 f(\omega) \coloneqq \sigma^2(\omega) + \phi \left(\mu(\omega) - \hat{\epsilon}\right)^2 + \psi \left(\sum_{i = 1}^{n} \omega_{i} - 1\right)^2,
\end{equation}
where $\omega \in [0,1]^n$ are the asset weights and $\phi, \psi \in \mathbb{R}_{\geq 0}$ are penalty factors that penalize violations of the return constraint~\eqref{eq:minreturn_constraint} and the normalization constraint~\eqref{eq:normalization_constraint}, respectively.
Here, we replace the ``$\geq$'' in the return constraint ~\eqref{eq:minreturn_constraint} by ``$=$''.
According to Markowitz portfolio theory the variance-minimal portfolio for $\mu(\omega) \geq$ $\xi$ has exactly return $\mu(\omega) = $$\xi$ for a sufficiently large $\xi$.
We elaborate on this in appendix~\ref{sec:efficient_frontier}.
Also, we select the return target as $\hat\epsilon=\epsilon+1\%$, as sampling-based methods return a distribution of portfolios whose returns scatter around the penalised target.
Since a feasibility analysis accepts any solution satisfying the original inequality $\mu(\omega) \geq \epsilon$, shifting the penalised target slightly upwards increases the number of samples that are feasible for the original problem.
The same target $\hat\epsilon$ is used for all QUBO-based methods.
Thus, the comparison between the solving methods in the following chapter remains unaffected.
From a theoretical point of view, finding optimal penalty factors is itself an NP-hard problem~\cite{alessandroni_alleviating_2025}.
Consequently, we employ a heuristic penalty approach in this work, which we detail in section~\ref{sec:benchmark_study_results}.

In order to derive a valid QUBO model,
the continuous variables $\omega_i \in [0,1]$ in~\eqref{eq:qubo_obj} have to be transformed to binaries.
As proposed in~\cite{sakuler_real_2023, palmer_quantum_2021}, we write
the portfolio weights $\omega_i \in [0,1]$ as a linear combination of
binary variables $\omega_{ij}\in\{0,1\}$, $j\in\{1,\dots,d+1\}$, via
\begin{equation}\label{eq:discretize}
 \omega_{i} = u_i \cdot \left( \frac{1}{2^d} \omega_{i, d+1} + \sum_{j=1}^{d} \frac{1}{2^{j}}\omega_{i, j} \right) \quad \forall i \in \{1,...,n\}.
\end{equation}
Here, $d\in \mathbb{N}$ controls the precision of discretization.
An advantage of this discretization is the implicit incorporation of the upper bounds $u_i$ on the asset weights.
Inserting~\eqref{eq:discretize} in~\eqref{eq:qubo_obj} yields the QUBO model for~\hyperlink{eq:min_vola}{\text{MinVola}}.

\section{Methods}\label{sec:methods}
In this section, we introduce the solution methods that we compare in our benchmark.
Furthermore, we explain how associated parameters are chosen.
For all methods, we leave parameters to default values if not stated otherwise.

\paragraph{Exact mathematical optimization solver.}
In this benchmark, we chose Gurobi~\cite{noauthor_leader_nodate} as the classical solver that provably solves the problem to optimality.
In order to ensure fairness when comparing the other methods to the exact solver, we solve the discretized version of the formulation~\hyperlink{eq:min_vola}{\text{MinVola}}.

\paragraph{Quantum annealing.}
Quantum annealing is a heuristic algorithm that runs on analog quantum computers~\cite{hauke_perspectives_2020,albash_adiabatic_2018}.
The concept of quantum annealing is based on the adiabatic theorem of quantum mechanics~\cite{messiah_quantum_1961}.
Quantum annealing prepares a quantum mechanical system in the ground state of some initial Hamiltonian $H_0$.
Then, it evolves the system according to
 \begin{equation}
 H(t) = A(t) H_0 + B(t) H_1 \,.
 \end{equation}
Here, $H_1$ is the problem Hamiltonian whose ground state encodes the solution to some optimization problem.
$A:[0,\tau]\rightarrow[0,1]$ is chosen such that $A(0) = 1$ and $A(\tau)=0$, where $\tau>0$ is called \emph{annealing time}.
Similarly, $B:[0,\tau]\rightarrow[0,1]$ is chosen such that $B(0) = 0$ and $B(\tau ) = 1$.
Now, the adiabatic theorem states that the system is in the ground state of $H_1$ at $t=\tau$ for $\tau \gg 1$
if the energy gap between the ground state and the first excited state is non-zero for all $t\in [0,\tau]$
and if $\partial_tH(t)$ is finite for all $t\in [0,\tau]$.

In our computational experiments between the months of June and September of 2025, we use a D-Wave Advantage 2 processor which has more than 4,400 physical qubits, arranged in the Zephyr topology~\cite{noauthor_advantage_nodate}.
However, the limited device connectivity requires an embedding of each QUBO variable into possibly multiple physical qubits~\cite{choi_minor-embedding_2008, choi_minor-embedding_2011}.
This embedding overhead increases with the density of the QUBO problem.
All considered QUBO instances have a density of 100\%.
For embedding, we therefore use the \texttt{DWaveCliqueSampler}~\cite{noauthor_samplers_nodate} from the python package \textit{dwave-system}~\cite{noauthor_dwave_system_2025}, which employs pre-defined embeddings for fully connected QUBO problems
and is thus more suited for dense problems than constructive embedding algorithms like the default \texttt{MinorMiner}~\cite{cai_practical_2014}.
Also, we test annealing times of $1$ µs, $5$ µs, $20$ µs and $50$ µs.
The \emph{annealing schedule}, which defines $A(t)$ and $B(t)$, is kept as default.
However, we vary the \emph{chain strength} parameter which can prevent the breaking of a chain of connected qubits representing a single QUBO variable.
We explicitly decided against more sophisticated quantum annealing concepts like warm-starts~\cite{schlütter2025hotstartingquantumportfoliooptimization} or special annealing schedules~\cite{Khezri_2022, tang_comparative_2024, venturelli_reverse_2019} because our goal is to test the quality of solutions returned by the basic provided infrastructure without focusing on presolving and fine-tuning the methodology and the associated parameters.

\paragraph{QAOA.} \label{sec:qaoa}
The quantum approximate optimization algorithm is a quantum-classical hybrid algorithm, originally proposed by Farhi et al.~\cite{farhi_quantum_2014}.
Like quantum annealing, QAOA is conceptually based on the adiabatic theorem of quantum mechanics~\cite{messiah_quantum_1961}.
However, in contrast to quantum annealing, QAOA is designed for gate-based, universal quantum computers.
QAOA depends on real-valued parameters $\gamma=(\gamma_1,...,\gamma_p)$ and $\beta=(\beta_1,...,\beta_p)$.
The hyper-parameter $p\in\mathbb{N}$ controls the complexity of the algorithm.
QAOA starts with the uniform superposition quantum state $\ket{+}^N$ and evolves it according to
 \begin{equation}
 \ket{\psi(\beta, \gamma)} = e^{-i \beta_{p} H_{M}} e^{-i \gamma_{p} H_{C}} \ ... \ e^{-i \beta_{1} H_{M}} e^{-i \gamma_{1} H_{C}} \ket{+}^N.
 \label{QAOA equation}
 \end{equation}
Here $H_{C}$ is the so-called \emph{problem Hamiltonian}, defined by
 \begin{equation}
 H_{C}\ket{x} = C(x) \ket{x} \quad \forall x \in \{0,1\}^N
 \end{equation}
where $C(x)$ is the QUBO cost function.
$H_{M}$ is called the \emph{mixing Hamiltonian} and is defined by
 \begin{equation}
 H_{M} = \sum_{i=1}^{N} X_{i}
 \end{equation}
where $X_{i}$ is the Pauli $X$ gate acting on the $i$-th qubit.
For given parameters $\beta$ and $\gamma$,
solutions to the QUBO problem are sampled from the quantum circuit implementing~\eqref{QAOA equation}.
In this work, we employ three different methods to compute the parameters $\beta$ and $\gamma$.
First, for $p=1$, we perform a grid search by analytically calculating the expectation value of the resulting QAOA circuit via the formula developed in~\cite{ozaeta_expectation_2022}.
We choose the parameter values that result in the minimum expectation value.
Second, we use the linear ramp QAOA (LR-QAOA) protocol introduced in~\cite{montanez-barrera_towards_2025}.
Accordingly, we set $\beta_i = (1-\frac{i}{p})\Delta_{\beta}$ and
$\gamma_i = (\frac{i+1}{p})\Delta_{\gamma}$ for $i\in\{1,\dots,p\}$.
In our experiments, we choose $\Delta_{\beta} = 0.3$ and $\Delta_{\gamma}=0.6$ as proposed in~\cite{montanez-barrera_towards_2025}.
Finally, we optimize the parameters with the local optimizer COBYLA~\cite{noauthor_minimizemethodcobyla_nodate, noauthor_scipy_2025} using a noiseless quantum simulator~\cite{noauthor_aersimulator_nodate, noauthor_qiskit_aer_2025}.
Here, we initialize the parameters with the LR-QAOA-formula.

After selecting parameters, we execute QAOA on the gate-based quantum computer \emph{ibm\_strasbourg} based on a 127 qubit \emph{Eagle} processor~\cite{abughanem_ibm_2025}.
The experiments on \emph{ibm\_strasbourg} were conducted in July 2025.
Circuits were transpiled by \textit{generate\_preset\_pass\_manager}~\cite{noauthor_generate_preset_pass_manager_nodate} with \textit{optimization\_level}$=1$.
No particular error mitigation techniques are set.
We also run a benchmark with an ideal quantum simulator~\cite{noauthor_aersimulator_nodate}.
Exponentially increasing memory requirements limit the classical simulation of QAOA to roughly 30 variables.
We note that our method of calculating parameters has no optimality guarantee and that better QAOA parameters might exist.
Finally, we remark that like for quantum annealing, we decided against more advanced concepts like warm-starts~\cite{Egger_2021} or constraint-preserving mixers~\cite{Scursulim_2026} since we aim at testing the existing quantum architecture without method fine-tuning.

\paragraph{Steepest Descent.}
The steepest descent local search tries to improve the current solution by investigating its neighborhood.
Here, the neighborhood is defined as all solutions having a Hamming distance of one to the current solution, which is equivalent to one-bit-flips.
If the best solution in the neighborhood improves the objective value, it is accepted as the new current solution.
Otherwise, the algorithm terminates.
The main advantage of the steepest descent search is its fast runtime.
A disadvantage is its inability to escape local minima.
In our benchmark study, we employ the open-source implementation available in~\cite{noauthor_dwavesystemsdwave-greedy_2024, noauthor_dwave_greedy_2022}.

\paragraph{Simulated Annealing.}
Simulated annealing is a popular classical heuristic.
The name of this algorithm is based on its similarity to the process of annealing in metallurgy,
which aims to harden a metal by controlled heating and cooling.
In its general form, simulated annealing starts with a random solution.
Then, simulated annealing iteratively considers a random solution in the neighborhood of the current solutions.
If this solution improves upon the current objective value, it is accepted as the new current solution.
Otherwise, it is still accepted as the new current solution with a probability decreasing with the number of iterations.
By this methodology, the algorithm explores a large fraction of the solution space at the start, while at the end,
objective-improving solutions are favored.
The algorithm returns the solution with the best objective value~\cite{Kirkpatrick1983}.
Several parameters can be modified, for example, the definition of the neighborhood function and the acceptance probability function.
In our computational study, we use an open-source implementation of simulated annealing~\cite{noauthor_dwave-samplers_nodate, noauthor_dwave_samplers_2025}.

\paragraph{Tabu Search.}
Tabu search is a classical heuristic conceptually similar to steepest descent.
Tabu search starts with a random solution, then explores the neighborhood of this solution and chooses the best solution in this neighborhood
while putting the previous solution on a tabu list.
Solutions on the tabu list are excluded from any neighborhood.
The tabu list has a limited length and solutions are removed in a first-in-first-out manner.
The tabu list allows the algorithm to escape local minima.
Termination criteria can be runtime, number of steps or objective value.
Adjustable parameters are the selection of the neighborhood function, the termination criterion and the length of the tabu list.
In our computational study we use an open-source implementation of tabu search~\cite{noauthor_dwavesystemsdwave-tabu_2024, noauthor_dwave_tabu_2022}.

\paragraph{Problem-specific heuristic.}
We also developed a problem-specific heuristic for the \hyperlink{eq:min_vola}{\text{MinVola}} version of the portfolio optimization problem which  does not require building the mathematical model.
For an instance with $n$ assets, it creates up to $n$ feasible solutions, but does not guarantee feasibility.
The heuristic starts with an empty portfolio and then increases the asset weight $\omega_i$ of some asset $i\in \{1,\dots,n\}$ by a fixed amount of $\delta >0$.
Then, it iteratively adds weight $\delta$ to the asset that reduces the portfolio variance the most while still satisfying constraint \eqref{eq:minreturn_constraint}.
In our case, we use $\delta = \frac{u_i}{2^d}$ to apply the same discretization as for the other methods.
If the return constraint can not be satisfied, the algorithm increases the weight of the asset with the largest return, i.e., it moves towards return feasibility rather than terminating.
The heuristic stops when the sum of weights is greater or equal to 1.
Finally, the weights are normalized and the solution is checked for feasibility
For a more detailed description and pseudocode of the problem specific heuristic, we refer to appendix~\ref{sec:minvola_heuristic}.

\section{Computational Experiments}\label{sec:benchmark_study_results}
In this section, we benchmark the solution methods discussed in Section~\ref{sec:methods} on the portfolio optimization problem test set introduced in Section~\ref{sec:original_problem}.
The goal of our computational experiments is to quantify the potential of currently available quantum computers for portfolio optimization.
Our study follows the general rules on good-practice for benchmarks from~\cite{acuaviva_benchmarking_2026}, which are relevance, reproducibility, fairness, verifiability and usability.
Our code and data are publicly available at~\cite{noauthor_stopferericportfolio_opt_benchmark_nodate}.

\paragraph{Benchmark procedure.}
The benchmark study was conducted on a standard notebook equipped with a 10-core, $16$ GB RAM Intel Core i7 CPU.
We compare the solution methods introduced in Section~\ref{sec:methods}, that is, QAOA, quantum annealing, simulated annealing, steepest descent, tabu search and the problem-specific heuristic on the \hyperlink{eq:min_vola}{\text{MinVola}} problem variant.

In the QUBO model~\eqref{eq:qubo_obj}, we set the penalty factors $\phi=\psi=1000$ for violations of the return constraint~\eqref{eq:minreturn_constraint} and the normalization constraint~\eqref{eq:normalization_constraint}, respectively.
We determine these factors by analyzing the magnitude of the objective and constraint terms.
We detail the derivation of the penalty factors in appendix \ref{sec:penalty_factors}, where we conduct a magnitude-analysis of the objective terms for the optimization problem.

With these penalty factors, a small absolute constraint violation of $0.01$ in the constraints~\eqref{eq:minreturn_constraint} or~\eqref{eq:normalization_constraint}
will lead to a penalization of $1000 \cdot 0.01^2 = 0.1$ which is in the same order of magnitude as the objective value.

Additionally, we choose the discretization coarseness in~\eqref{eq:discretize} as $d=3$.
This leads to $d+1=4$ times as many variables in the discretized model as in the continuous model.
Consequently, for an instance with $n$ assets, the resulting QUBO has $4n$ variables.
The value of $d=3$ results from a trade-off between QUBO size and discretization precision.

To ensure fairness among the methods, we enforce a time limit of 60 seconds on the execution time of the algorithm on the device, i.e., the sampling time.
We do not include the model construction into the time limit since the time for model construction varies depending on which library is used (like qiskit\_optimization or dimod).
Moreover, our process of first building the Quadratic Program and then transforming it into a QUBO could be sped up by directly building the QUBO, possibly in a fast programming language like C++.
For the same reason, we do not include the result analysis time, i.e., calculating the QUBO objective values, which could be sped up by a faster implementation.
We also do not include the embedding time into the time limit since the problem is always fully connected and thus, ideally, the embedding for a problem is calculated just once and reused for every other problem of the same size.
Also, cloud connection, job submission and queueing times are not considered because we are interested in analyzing the capabilities of the quantum computer itself rather than access overhead.
For the grid-search QAOA, we split up the 60 seconds into 30 seconds of training and 30 seconds of sampling with the best found parameters.
We configure the size of the grid such that its execution time takes roughly 30 seconds.
For the QAOA with optimized parameters, we spent a maximum of 52 seconds on training while the rest is used to exclusively sample with the best found parameters.
This ensures that we have at least a minimum amount of time for sampling with high-quality parameters.
Of course, the sampling results found during the training of QAOA are also considered in the final result.
We remark that the accounting of runtime is deliberately favourable to the quantum methods since there is some infrastructure overhead involved like cloud connection, job submission and queueing.
Finally, we note that a slightly smaller budget than $60$~seconds was configured in the implementation as a safety margin to ensure that the limit is never exceeded.

After sampling within the boundaries of the time limit, we calculate the QUBO objective values of the entire sample. 
We take the 100 best solutions and execute a postprocessing step by scaling the weights such that they sum up to 1 in order to satisfy the normalization constraint~\eqref{eq:normalization_constraint}.
Then, we analyze these solutions with respect to feasibility and objective value for the original \hyperlink{eq:min_vola}{\text{MinVola}} problem.

We then return the best found solution.
This resembles a practical setting where one is usually interested in the best solution rather than
statistical performance measures like expectation values or quantiles.

As our primary solution quality metric, we consider the \emph{approximation ratio} which is defined as
 \begin{equation}
 \Theta \coloneqq \frac{f_m}{f_\text{opt}}\geq 1.
 \end{equation}
Here $ f_m $ is the objective value returned by the method ${m}$, and $ f_\text{opt} $ is the value of the optimal solution of the discretized version of the problem.
which we calculate by an exact mathematical optimization solver.
We have $ f_m \geq f_\text{opt} \geq 0 $ since \hyperlink{eq:min_vola}{\text{MinVola}} is a minimization problem and the portfolio variance is always non-negative.
Thus, a value of $\Theta$ close to $1$ indicates near-optimal solutions.

In addition, we measure the \emph{feasibility percentage}, which we define as the fraction of the 10 test instances per problem size for which the considered method returns at least a single feasible solution.

Finally, the \emph{number of samples} measures how many samples are generated by the respective algorithm within the time limit.

\begin{figure}[p]
 \centering
 \begin{subfigure}[b]{0.9\textwidth}
 \includegraphics[width=\linewidth]{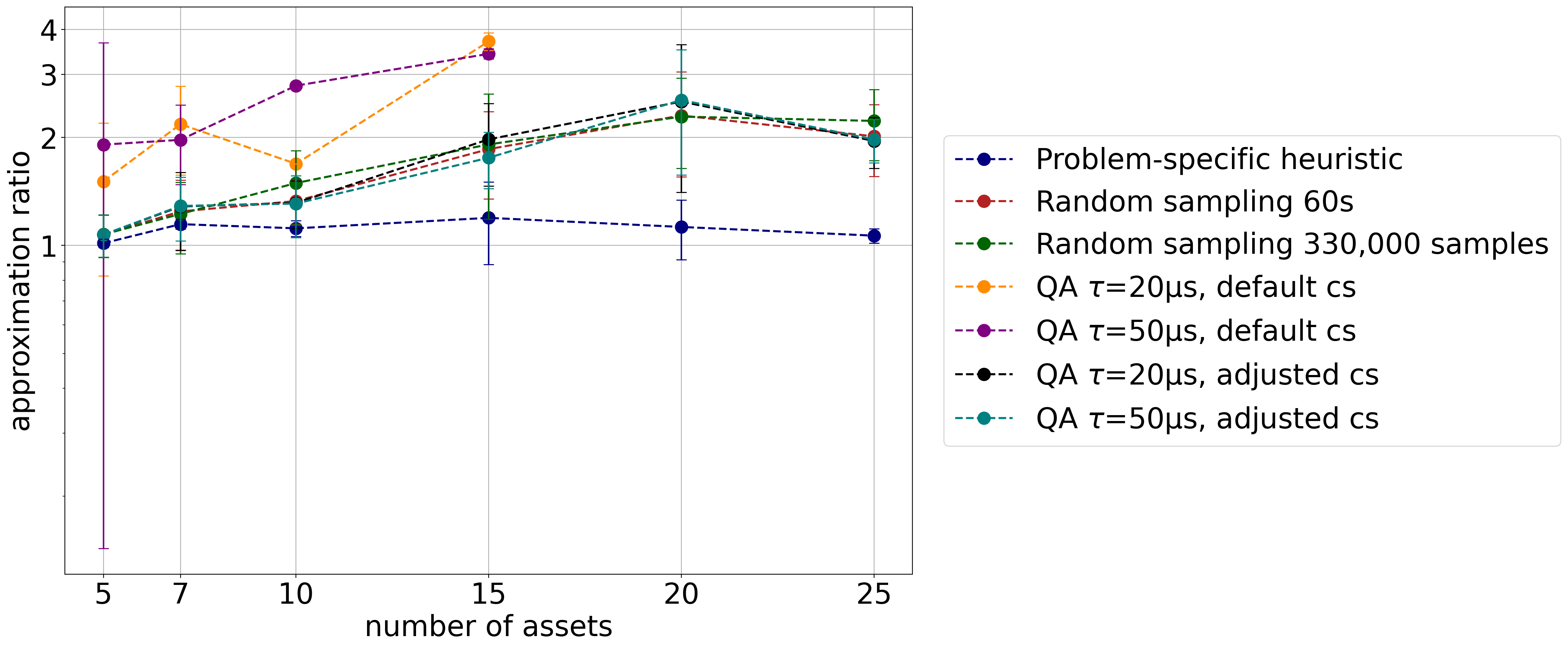}
 \caption{quantum annealing approximation ratio}
 \label{fig:annealing_approximation_ratio}
 \end{subfigure}
 \begin{subfigure}[b]{0.9\textwidth}
 \includegraphics[width=\linewidth]{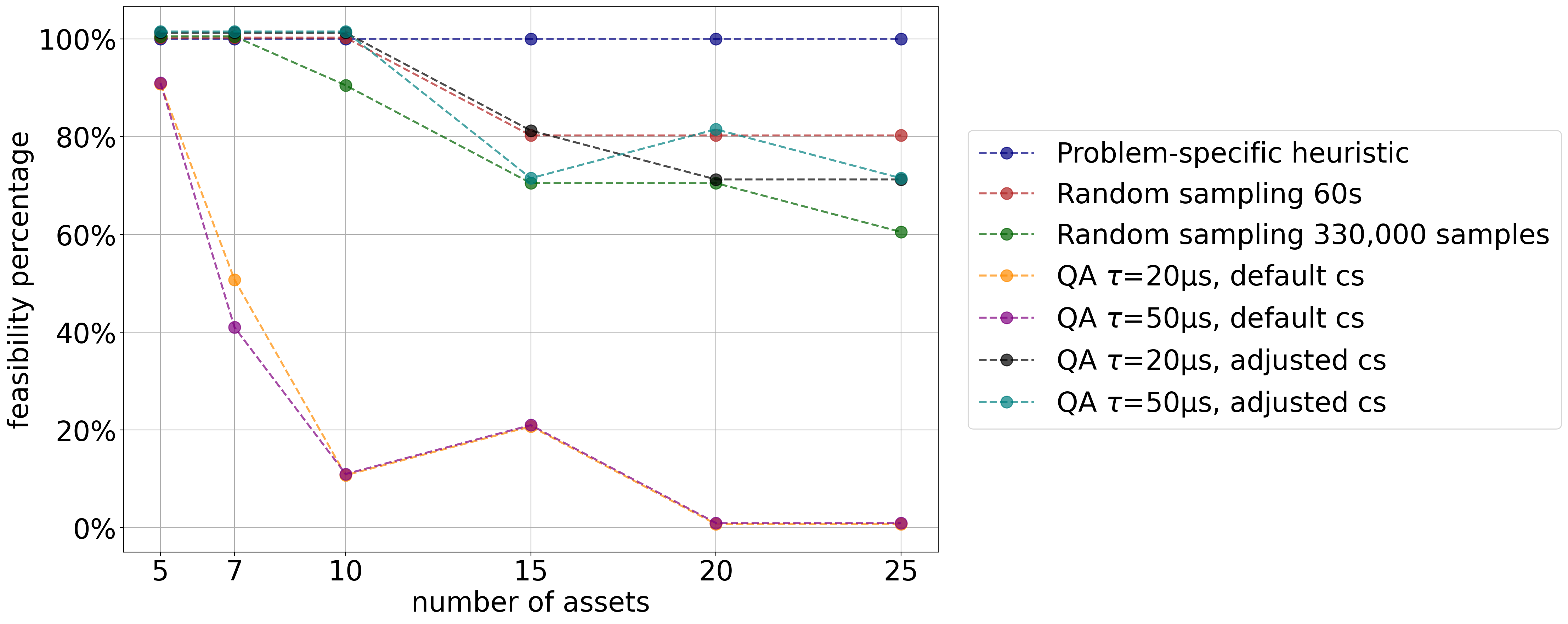}
 \caption{quantum annealing feasibility percentage}
 \label{fig:annealing_feasibility_percentage}
 \end{subfigure}
 \begin{subfigure}[b]{0.9\textwidth}
 \includegraphics[width=\linewidth]{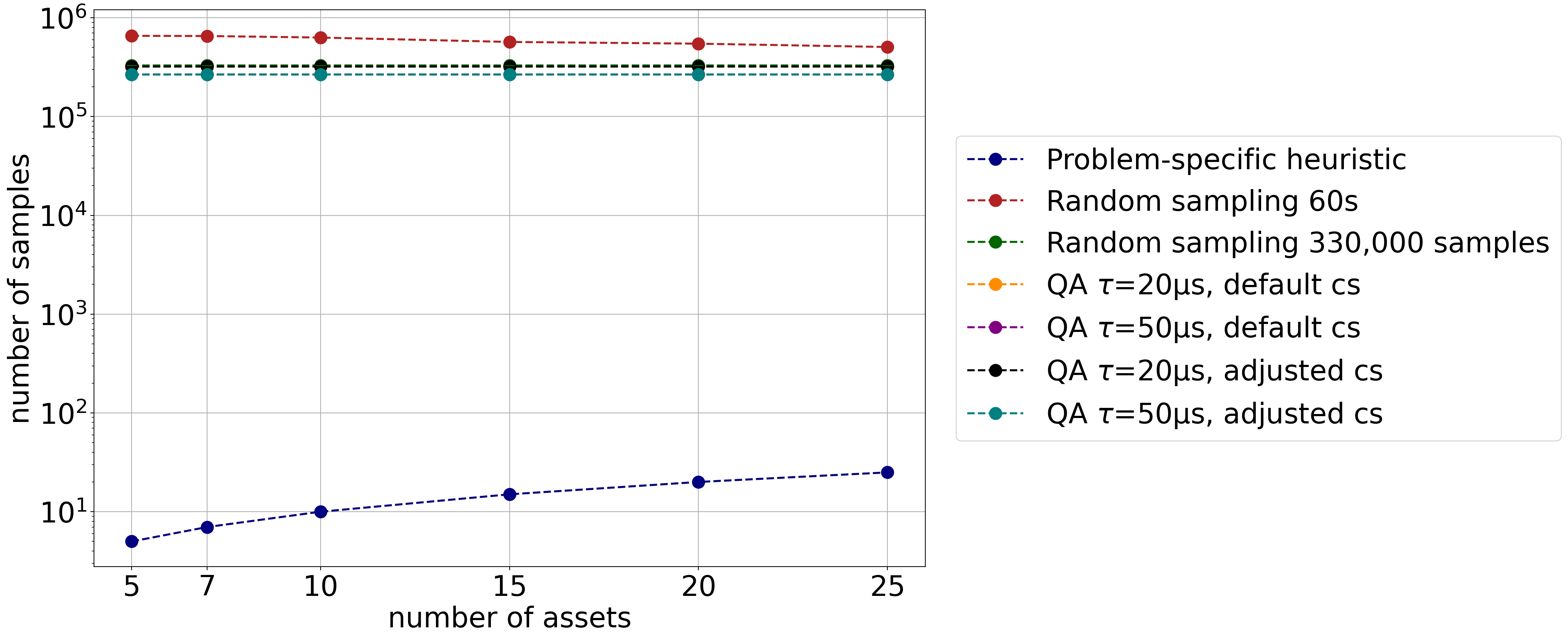}
 \caption{quantum annealing number of samples}
 \label{fig:annealing_number_of_shots}
 \end{subfigure}
 \label{fig:annealing_results}
 \caption{
 We report the approximation ratio, feasibility percentage and number of samples for quantum annealing.
 We compare against the problem-specific heuristic and random sampling as baselines.
 Data points are averages over $10$ instances and error bars show the empirical standard deviation.
 Additionally, distributions of the approximation ratios over the $10$ instances can be viewed in the appendix chapter \ref{sec:approximationratio_boxplots}.
 Missing data points in the approximation ratio are due to the absence of feasible solutions.
 }
\end{figure}

\paragraph{Benchmark Results.}
We first discuss the results of quantum annealing, shown in Figures~\ref{fig:annealing_approximation_ratio},~\ref{fig:annealing_feasibility_percentage} and~\ref{fig:annealing_number_of_shots}.
We observe that quantum annealing with the default calculation of the chain strength cs yields the worst approximation ratio (Figure~\ref{fig:annealing_approximation_ratio}) and the lowest feasibility percentage (Figure~\ref{fig:annealing_feasibility_percentage}) in the red and pink lines.
In particular, even random sampling has a lower average approximation ratio on the same number of samples while finding more feasible solutions (compare green and yellow lines in Figures~\ref{fig:annealing_approximation_ratio},~\ref{fig:annealing_feasibility_percentage},~\ref{fig:annealing_number_of_shots}).

After observing high chain break fractions, we tested different chain strength parameters.
Since all QUBO instances of equal size are fully connected and embedded using the same clique-embedding strategy, the resulting embeddings exhibit identical chain structures and chain lengths across instances of the same size.
Empirically, the observed chain break fractions were nearly identical across different problem instances of equal size, indicating that chain break behavior is dominated primarily by problem size and embedding geometry rather than by instance-specific QUBO coefficients.
For problems of $5,7,10$ and $15$ assets we set $cs=3$, for $20$ assets we set $cs=4$ and for $25$ assets we set $cs=5$.
These parameters result in chain break fractions in the desired range of $0\%$ to $5\%$ as recommended by D-wave during a private correspondence.
We would like to point out that such chain strength tuning can only improve the quantum annealing results, but it adds a bias to the reported performance estimate.
The results of quantum annealing with the customized chain strengths are shown in the black and gray lines in Figures~\ref{fig:annealing_approximation_ratio},~\ref{fig:annealing_feasibility_percentage} and~\ref{fig:annealing_number_of_shots}.
We observe that quantum annealing with adjusted chain strength is roughly on par with random sampling in terms of both feasibility percentage and approximation ratio.
Nevertheless, we observe in Figures~\ref{fig:annealing_approximation_ratio} and \ref{fig:annealing_feasibility_percentage} that quantum annealing performs significantly worse than the problem-specific heuristic.
We conjecture that the high density of the QUBO problem is a reason for the suboptimal performance of quantum annealing.
Embedding the required all-to-all connectivity leads to long chains of qubits for each variable.
Emphasizing the large embedding overhead, we note that
the largest problem size that can still be embedded on the quantum annealing processor with $4,400+$ physical qubits has 25 assets, which corresponds to 100 QUBO variables.
Moreover, varying the annealing time between $20 \ \mu s$ and $50 \ \mu s$ did not significantly improve the results, compare pink versus red and grey versus black in Figures~\ref{fig:annealing_approximation_ratio} and~\ref{fig:annealing_feasibility_percentage}.
We remark that we have conducted additional experiments with annealing times of $1 \ \mu s$ and $5 \ \mu s$.
Those results did not differ significantly from the results with annealing times of $20 \ \mu s$ and $50 \ \mu s$.
Finally, in Figure~\ref{fig:annealing_number_of_shots}, we observe that the number of samples generated in 60 s is independent of the problem size.
From this, we conclude that the runtime of quantum annealing does not significantly increase with problem size, which might lead to an advantage of future annealing hardware for very large instances.

\begin{figure}[p]
 \centering
 \begin{subfigure}[b]{0.9\textwidth}
 \includegraphics[width=\linewidth]{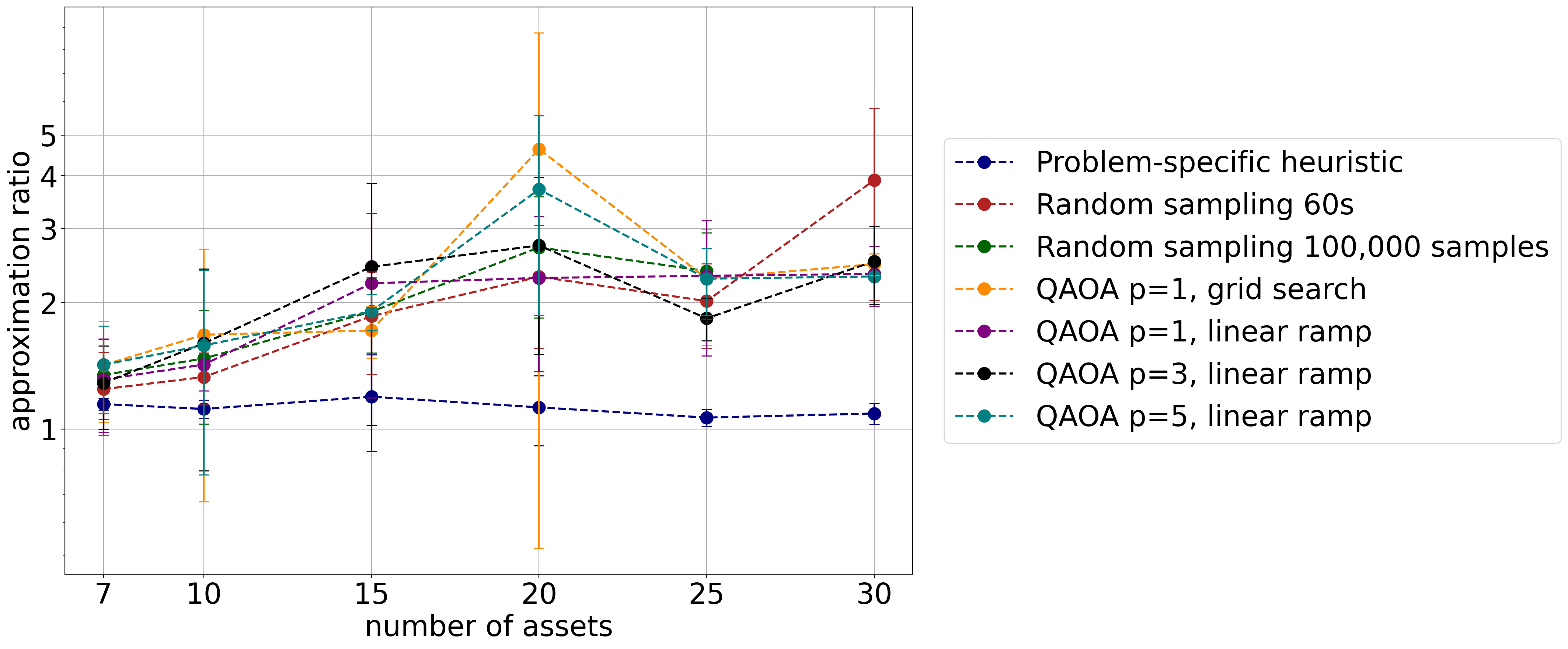}
 \caption{QAOA approximation ratio}
 \label{fig:qaoa_approximation_ratio}
 \end{subfigure}
 \begin{subfigure}[b]{0.9\textwidth}
 \includegraphics[width=\linewidth]{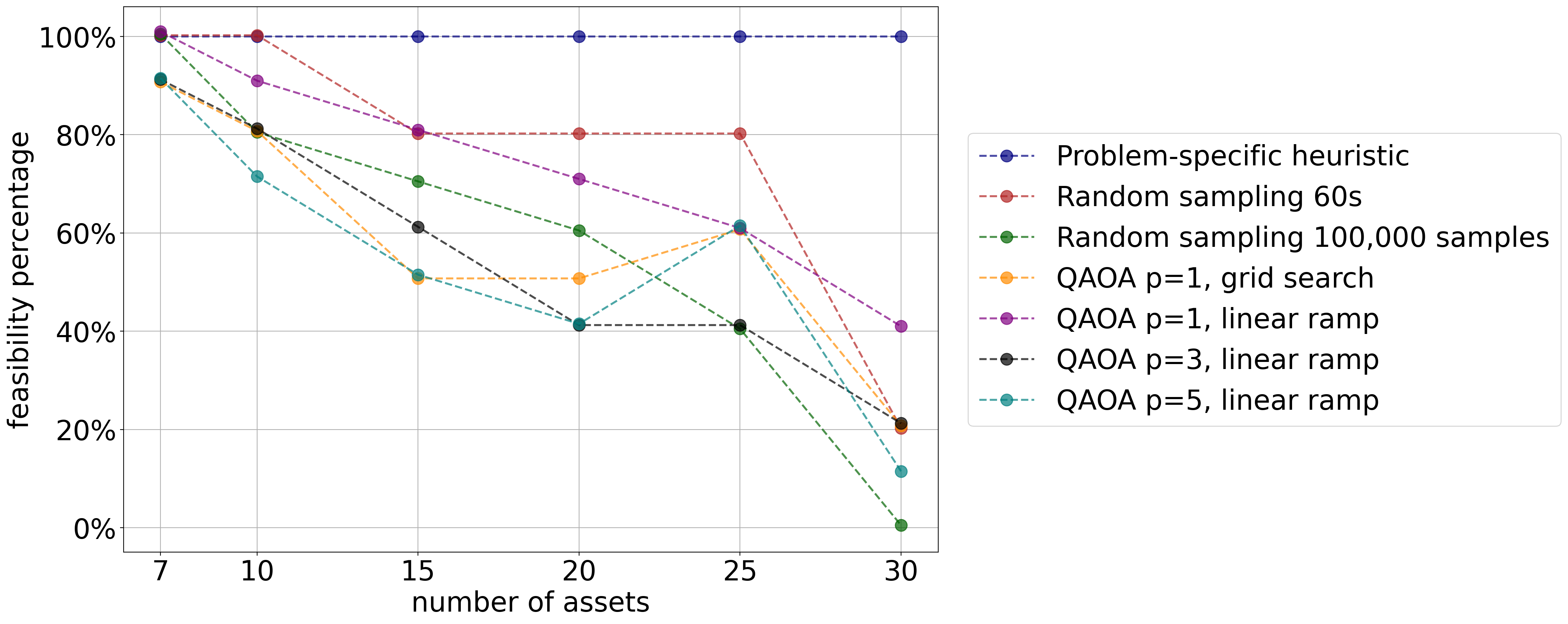}
 \caption{QAOA feasibility percentage}
 \label{fig:qaoa_feasibility_percentage}
 \end{subfigure}
 \begin{subfigure}[b]{0.9\textwidth}
 \includegraphics[width=\linewidth]{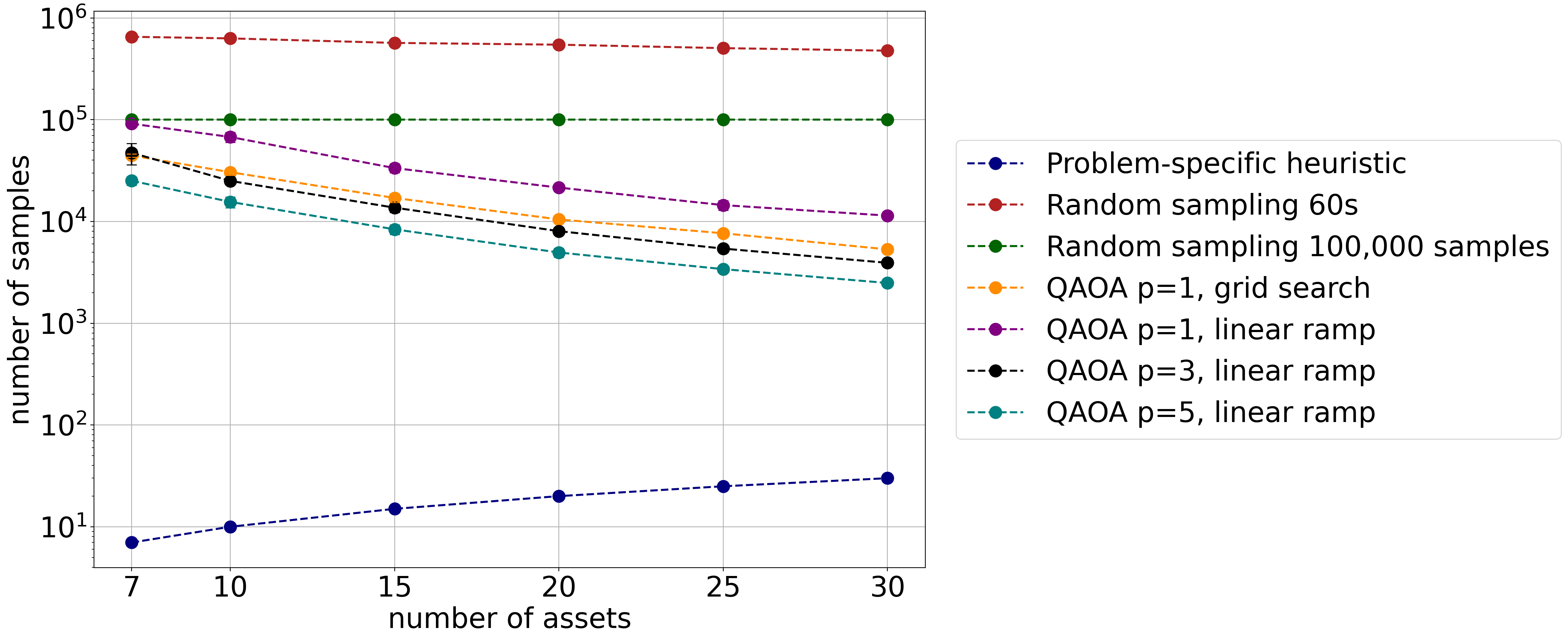}
 \caption{QAOA number of samples}
 \label{fig:qaoa_number_of_shots}
 \end{subfigure}
 \label{fig:qaoa_results}
 \caption{
 We report the approximation ratio, feasibility percentage and number of samples for QAOA.
 We compare against the problem-specific heuristic and random sampling as baselines.
 Data points are averages over $10$ instances and error bars show the empirical standard deviation.
 Additionally, distributions of the approximation ratios over the $10$ instances can be viewed in the appendix chapter \ref{sec:approximationratio_boxplots}.
 Missing data points in the approximation ratio are due to the absence of feasible solutions.
 }
\end{figure}

Next, we turn to the results for QAOA, shown in Figures~\ref{fig:qaoa_approximation_ratio},~\ref{fig:qaoa_feasibility_percentage},~\ref{fig:qaoa_number_of_shots}.
Similar to quantum annealing, QAOA often falls short on finding feasible solutions, which results in decaying feasibility percentages in Figure~\ref{fig:qaoa_feasibility_percentage}.
Even when QAOA returns feasible solutions, their approximation ratios quickly turn suboptimal when increasing the problem size.
We observe approximation values of $\geq 2$ for problems with $20$ or more assets in Figure~\ref{fig:qaoa_approximation_ratio}.
In general, QAOA does not perform significantly better than random sampling with respect to feasibility and approximation ratio.
Compared to the problem-specific heuristic, all QAOA configurations deliver lower feasibility percentages and worse approximation ratios.
The QAOA configuration with the best results is the 1-layer linear-ramp (red lines) which returns a lower approximation ratio with a higher feasibility percentage on a similar number of samples than the remaining QAOA configuration.
Moreover, we observe that increasing the number of layers worsens the performance although, in theory, the performance improves with the number of layers.
Other experiments with 2 and 4 layers affirm this observation while we do not show their results here for clarity.
We attribute this paradox mainly to two reasons.
First, more layers increase the circuit execution time.
Consequently, less shots can be executed in the same time, which can be seen in Figure~\ref{fig:qaoa_number_of_shots}.
Second, the effect of noise increases with the circuit size.
A reason for the decaying performance of QAOA with larger problem instances is the rapidly increasing number of swap gates that are required to transpile the QAOA circuit with all-to-all-connectivity to the quantum computer with limited connectivity.
For example, for 20 assets, the circuit depth of 1-layer QAOA increases from $161$ to $\approx 11,000$ during transpilation.

The results for the classically simulated QAOA are only reported in appendix \ref{sec:simqaoa_results}, as the largest instance which can still be simulated has only $7$ assets.
Finally, we remark that we did not run QAOA experiments in which we optimize the parameters on the hardware since the communication and queuing times are prohibitively large which is highly impractical for a parameter optimization loop.
We remark that IBM offers a session mode which reduces queuing, however, it drastically consumes more hardware access time than for the jobs themselves.

\begin{figure}[p]
 \centering
 \begin{subfigure}[b]{0.9\textwidth}
 \includegraphics[width=\linewidth]{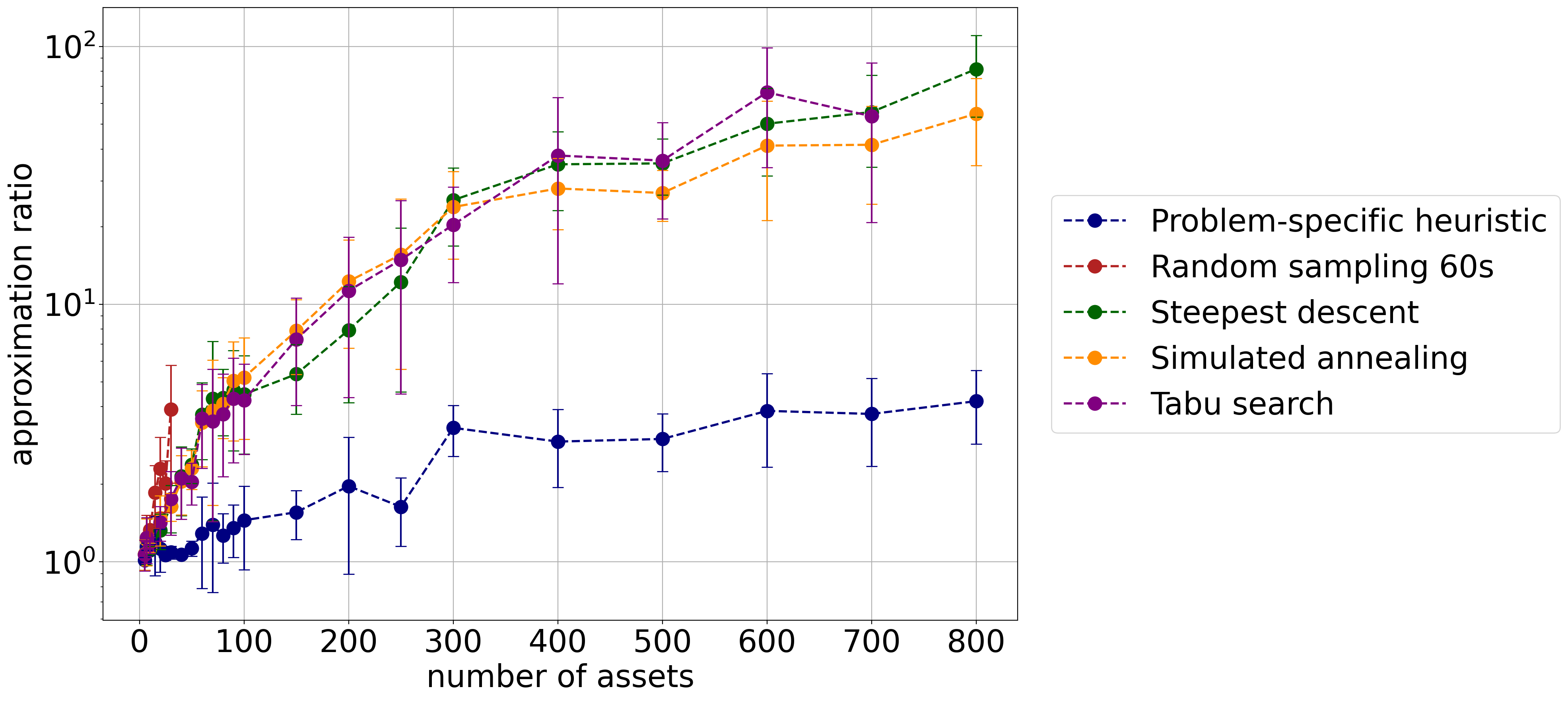}
 \caption{heuristics approximation ratio}
 \label{fig:heuristics_approximation_ratio}
 \end{subfigure}
 \begin{subfigure}[b]{0.9\textwidth}
 \includegraphics[width=\linewidth]{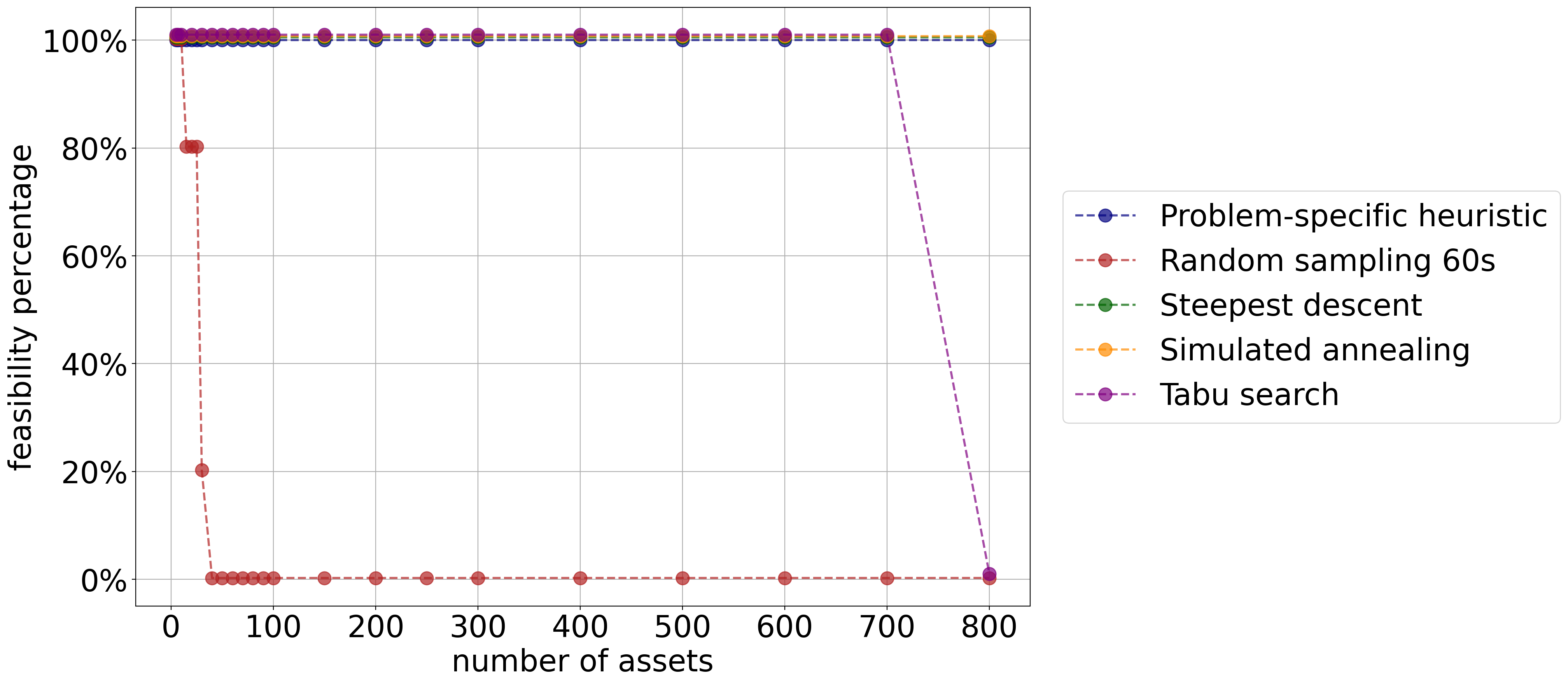}
 \caption{heuristics feasibility percentage}
 \label{fig:heuristics_feasibility_percentage}
 \end{subfigure}
 \begin{subfigure}[b]{0.9\textwidth}
 \includegraphics[width=\linewidth]{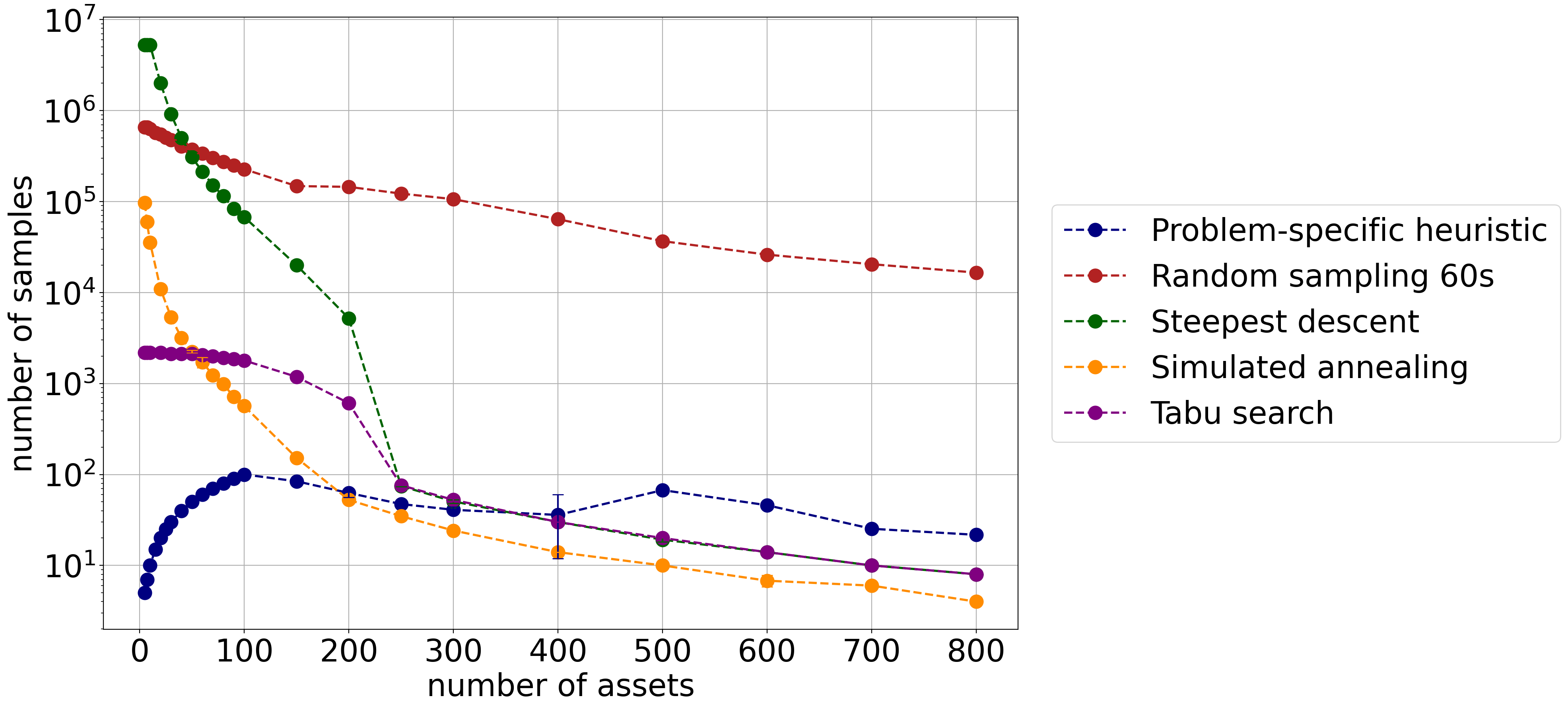}
 \caption{heuristics number of samples}
 \label{fig:heuristics_number_of_shots}
 \end{subfigure}
 \label{fig:heuristics_results}
 \caption{
 We report the approximation ratio, feasibility percentage and number of samples for the heuristics.
 We compare against the problem-specific heuristic and random sampling as baselines.
 Data points are averages over $10$ instances and error bars show the empirical standard deviation.
 Additionally, distributions of the approximation ratios over the $10$ instances can be viewed in the appendix chapter \ref{sec:approximationratio_boxplots}.
 Missing data points in the approximation ratio are due to the absence of feasible solutions.
 }
\end{figure}

Lastly, we compare the results for classical heuristics in Figures~\ref{fig:heuristics_approximation_ratio},~\ref{fig:heuristics_feasibility_percentage} and~\ref{fig:heuristics_number_of_shots}.
All open-source implementations (steepest descent, simulated annealing and tabu search) are executed with default settings.
In general, all heuristics are able to find feasible solutions for nearly all problem instances, see Figure~\ref{fig:heuristics_feasibility_percentage}.
Only tabu search does not return feasible solutions for all instances with 800 assets.
On the other hand, random sampling failed to return any feasible solutions starting from 40 assets.
The approximation ratios of simulated annealing, steepest descent and tabu search exceed the value of $2$ starting at 50 assets and the value of $10$ at 250 or more assets.
The problem-specific heuristic performs best with respect to the average approximation ratio.
The approximation ratios of the other heuristics are comparable despite differing significantly in the numbers of samples (Figure~\ref{fig:heuristics_number_of_shots}).
We remark that all heuristics rely on objective evaluation which can take a significant amount of time on large instances.
This explains the decreasing number of samples for larger problem instances.
 For example, for $500$ assets, the problem-specific heuristic is only able to return roughly $100$ out of $500$ possible solutions in $60$ s.
Thus, quantum methods might be beneficial for even higher asset numbers since they require classical objective evaluation only during the sample evaluation process and not during sampling itself.
Nevertheless, the performance of the problem-specific heuristic sets the bar high for a possible quantum advantage.

\begin{figure}[p]
 \centering
 \begin{subfigure}[b]{0.9\textwidth}
 \includegraphics[width=\linewidth]{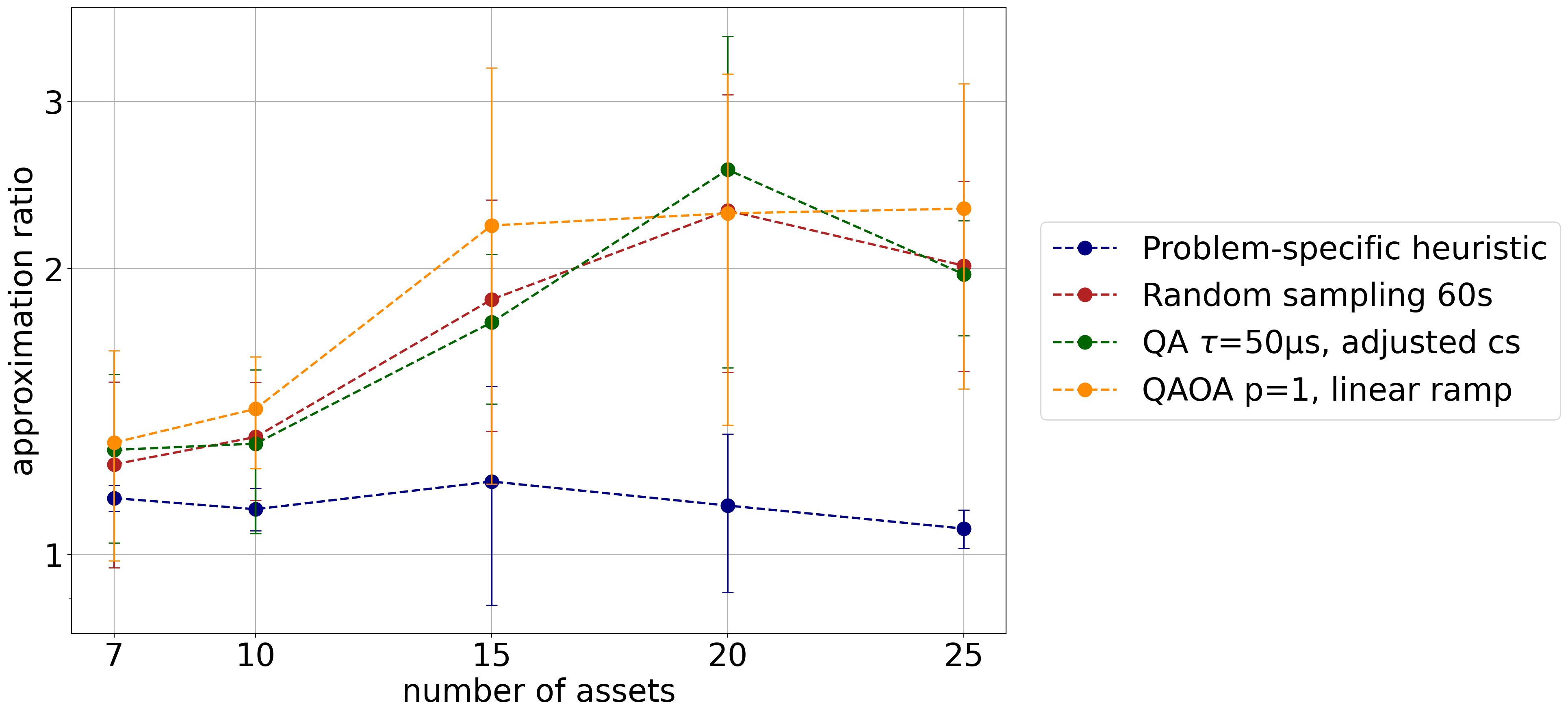}
 \caption{Best methods: approximation ratio}
 \label{fig:bestmethod_approximation_ratio}
 \end{subfigure}
 \begin{subfigure}[b]{0.9\textwidth}
 \includegraphics[width=\linewidth]{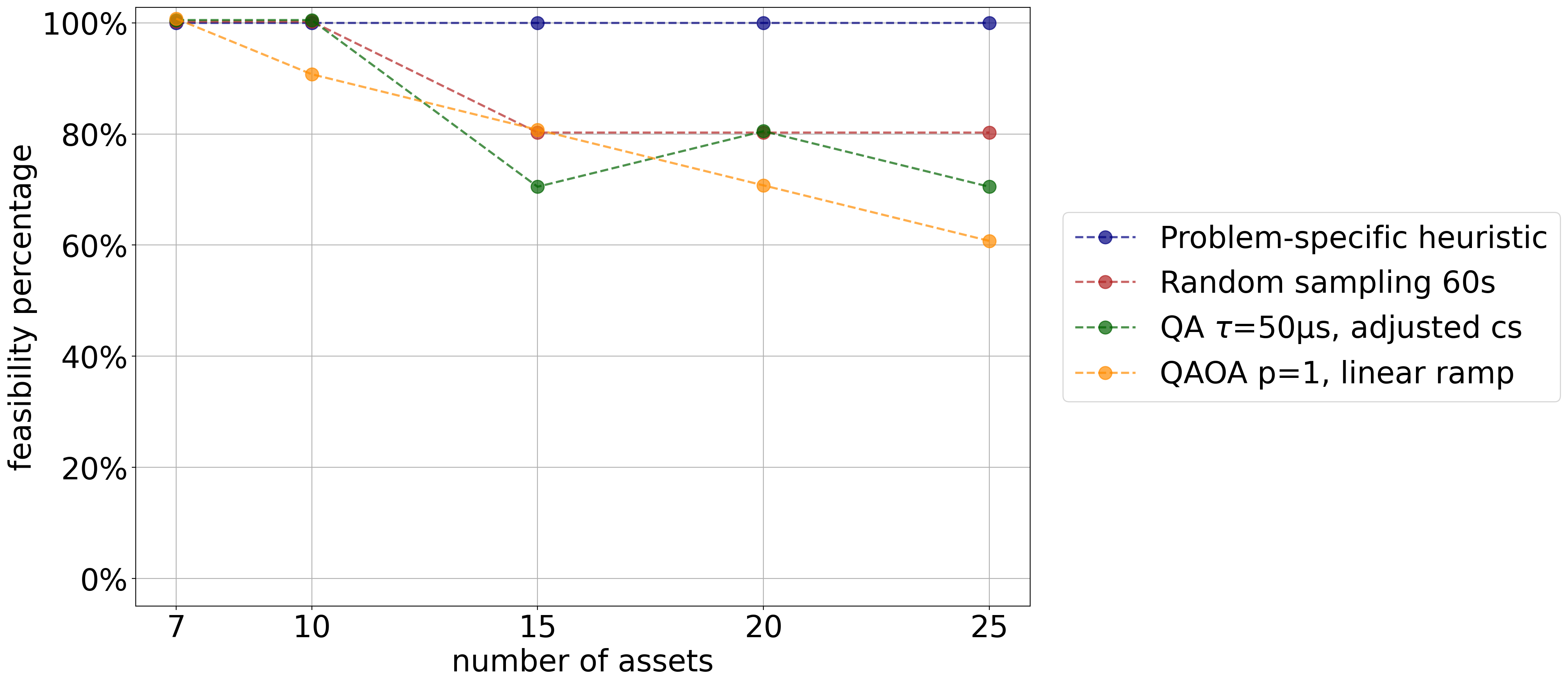}
 \caption{Best methods: feasibility percentage}
 \label{fig:bestmethod_feasibility_percentage}
 \end{subfigure}
 \begin{subfigure}[b]{0.9\textwidth}
 \includegraphics[width=\linewidth]{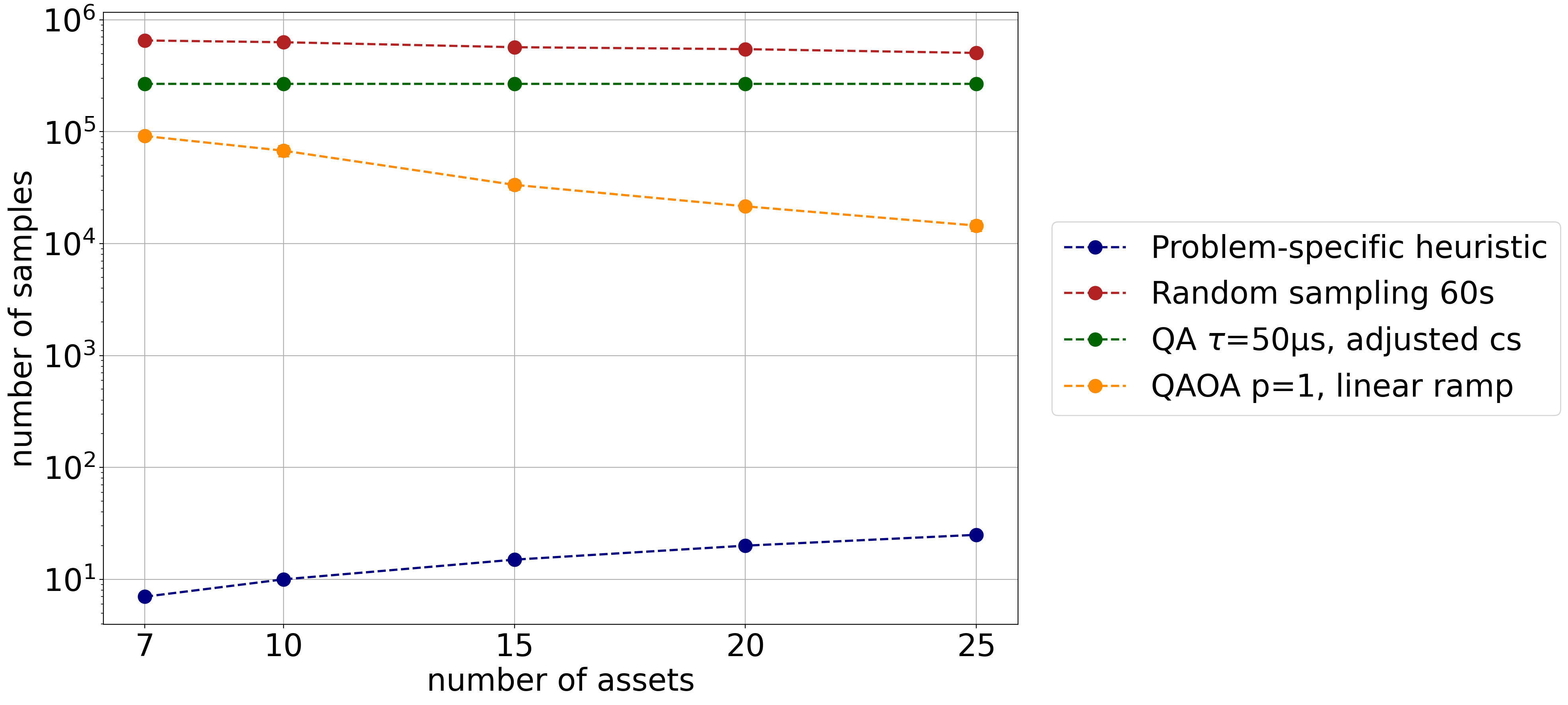}
 \caption{Best methods: number of samples}
 \label{fig:bestmethod_number_of_shots}
 \end{subfigure}
 \label{fig:bestmethod_results}
 \caption{
 Comparison of the best performing configurations from our computational study.
 The best performing configurations are the Problem-specific heuristic, quantum annealing with annealing time of $50\mu s$ and adjusted chain strengths as well as QAOA with 1 layer and linear-ramp parameters. For comparison purposes, random sampling for $60$ seconds is also visualized.
 We report the approximation ratio (a), the feasibility percentage (b) and the number of samples generated during the given time limit of $60$ seconds (c).
 Data points are averages over $10$ instances and error bars show the empirical standard deviation.
 }
\end{figure}

In Figures~\ref{fig:bestmethod_approximation_ratio},~\ref{fig:bestmethod_feasibility_percentage} and~\ref{fig:bestmethod_number_of_shots}, we compare the most promising approaches from quantum annealing, QAOA and classical heuristics in order to rank them.
In particular, we consider quantum annealing with $50$ µs annealing time and adjusted chain strength, the single-layer QAOA with linear ramp parameters and the problem-specific heuristic.

We observe that the problem-specific heuristic has a larger feasibility percentage and a better approximation ratio than the best configurations of both QAOA and quantum annealing.
Furthermore, both QAOA and quantum annealing perform roughly as good as random sampling.
Notably, random sampling generates significantly more solutions in 60 s (Figure~\ref{fig:bestmethod_number_of_shots}).
Finally, we conclude that quantum annealing slightly outperforms QAOA since quantum annealing generates more samples,
finds more feasible solutions
and those solutions are also of slightly higher quality.
The difference in the number of generated samples would get larger for increasing problem size since the QAOA circuit length and thus its execution time grows with the number of variables
whereas the annealing time stays constant.

\section{Conclusion}\label{sec:conclusion}
In this work, we conducted an extensive benchmark of quantum computing for portfolio optimization.
Recently, several works suggested portfolio optimization as a suitable candidate for a possible quantum advantage.
Our experiments, however, highlight the challenges of achieving such an advantage in practice.
We consider a variance-minimizing variant of portfolio optimization which we have shown to be more difficult for classical optimizers than return-maximizing or multi-objective formulations.
We then compared both classical and quantum methods on $270$ problem instances from real-world stock data.
In our benchmark, we impose a time limit of $60$ seconds on the sampling and result analysis to ensure fairness among the methods.
Our main conclusion is that classical heuristics like simulated annealing, steepest descent, tabu search and a problem-specific heuristic clearly outperform QAOA and quantum annealing regarding solution feasibility and quality.
Regarding classical exact optimizers, Gurobi and SCIP differ significantly in their solution times.
Here, Gurobi is more than $1000$ times faster for large problem instances, solving problems with 1000 assets in the order of seconds.
Comparing quantum methods, we observed that quantum annealing slightly outperformed the gate-based QAOA in terms of feasibility and quality.
However, this superiority of quantum annealing only appeared after fine-tuning chain strengths.
Because the adjusted chain strengths improve feasibility and approximation quality, any residual overfitting would bias the results in favor of quantum annealing and therefore does not affect the main conclusion of this study.
While being outperformed by classical heuristics, QAOA and quantum annealing also did not clearly differ from random sampling within the time limitation of 60 seconds.
Here, we emphasize that we observed a clear difference in the distribution of objective values when comparing QAOA and quantum annealing to random sampling.
However, in this work we considered only the best solution instead of statistical measures of the solution distribution.
We conjecture the all-to-all connectivity of the problem to be the main reason for the poor performance of QAOA and quantum annealing.
Dense problems generate a large qubit overhead during embedding and a large gate overhead during transpilation.
Finally, we remark that the considered problem variant is one of the simpler formulations of portfolio optimization.
For more complicated problem variants with additional constraints and variables, further experiments are required to study the performance of quantum methods.

\section*{Acknowledgements}
We thank D-Wave for their efforts in reviewing our experiments and for their recommendations on parameter settings.
This research was conducted within the
Bench-QC project, a lighthouse project of the Munich
Quantum Valley initiative, which is supported by the
Bavarian state with funds from the Hightech Agenda
Bayern Plus.

\clearpage
\appendix
\section*{Appendix}
\addcontentsline{toc}{section}{Appendix}

\section{Generating Stock Data}
\label{sec:generating_stock_data}
For our benchmark, we create portfolio problem instances from real world stock data.
We use the Python package yfinance~\cite{noauthor_yfinance_2025} to retrieve historical NASDAQ closing prices $p_{i,t}$ of asset $i$ at time $t$.
From the price data, we calculate the \emph{daily asset returns}
 \begin{equation}
 rd_{i,t} = \frac{p_{i,t}}{p_{i, t-1}} - 1 \quad \forall i \in N,\ t \in T
 \end{equation}
and the \emph{average daily asset returns}
 \begin{equation}
 rav_{i} = \frac{1}{|T|} \sum_{t \in T} rd_{i,t} \quad \forall i \in N.
 \end{equation}
For the portfolio optimization problem, we work with \emph{annualized asset returns}, which are defined by
 \begin{equation}
 r_{i} \coloneqq \left( \prod_{t \in T}(1+rd_{i,t}) \right) ^{\frac{252}{|T|}} - 1 \quad \forall i \in N.
 \end{equation}
The value of $252$ is the average yearly amount of business days at the stock exchange.
The estimated \emph{annualized covariances} for all asset combinations of asset $(i,j)$ are calculated by
 \begin{equation}\label{eq:def_sigma}
 \sigma_{ij} \coloneqq \frac{252}{|T|} \sum_{t \in T} (rd_{i,t} - rav_{i}) (rd_{j,t} - rav_{j}) \quad \forall i,j \in N.
 \end{equation}
The covariance matrix is used in its raw form without regularization because of the large number of observations in our test data set of $4$ years.
Now, the annualized asset returns $r_{i}$ and the annualized asset covariances $\sigma_{ij}$ are used to create instances of the portfolio optimization problem.

\section{Efficient Frontier}\label{sec:efficient_frontier}
\begin{figure}
 \centering
 \includegraphics[width=0.7\linewidth]{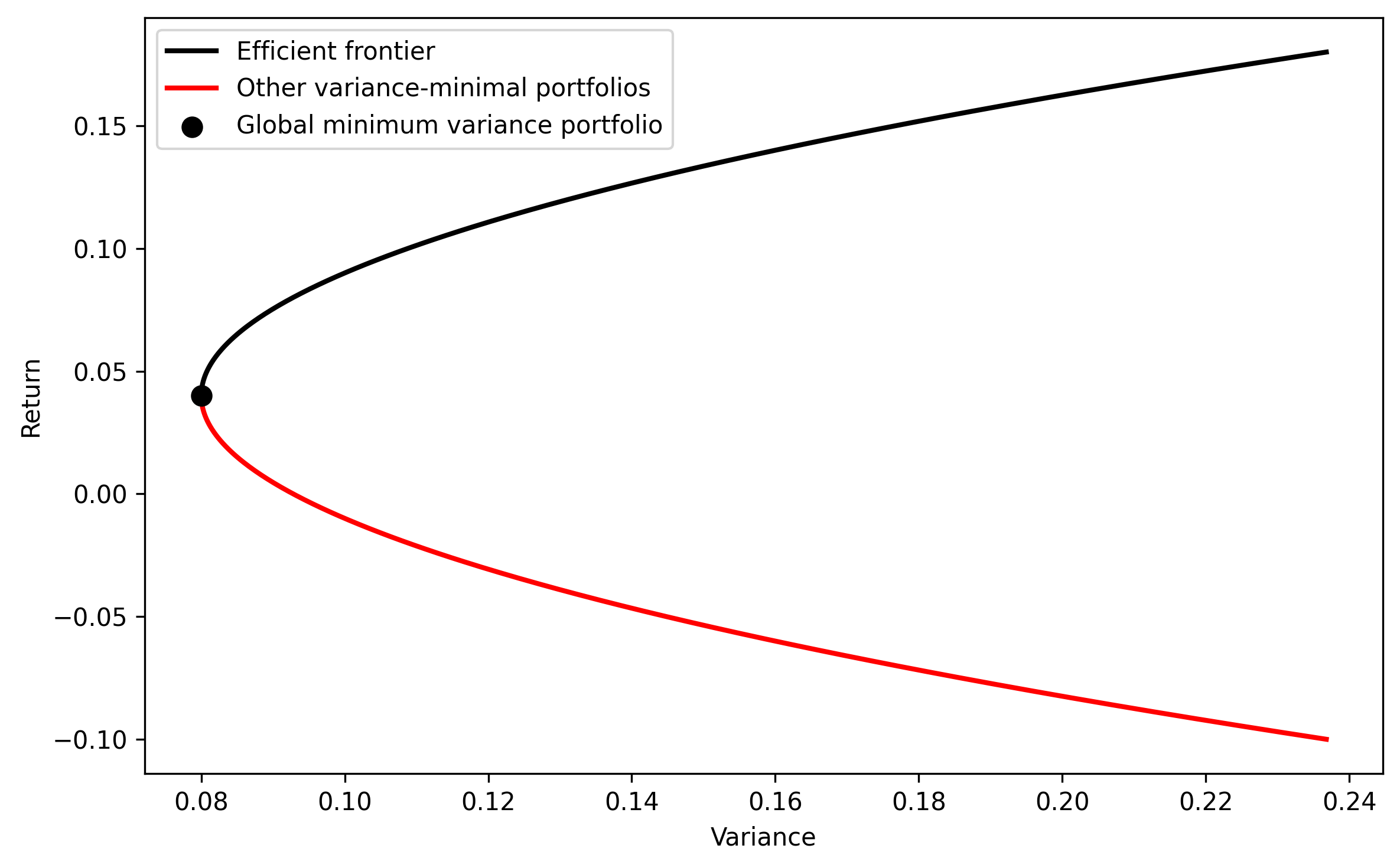}
 \caption{Visualization of the concept of the efficient frontier. The axes have arbitrary units.}
 \label{fig:efficient_frontier}
\end{figure}

Our argument for replacing the $\geq$ with $=$ in model~\hyperlink{eq:min_vola}{MinVola} relies on the \emph{efficient frontier} together with the selection of $\epsilon$.
According to the Markowitz Portfolio Theory~\cite{markowitz_portfolio_1952}, when displaying the minimum portfolio variance for given portfolio returns, a hyperbola arises~\cite{merton_efficient_frontier_1972}.
The upper part of this hyperbola is called the efficient frontier.
We visualize this concept by a schematic illustration in Fig.~\ref{fig:efficient_frontier}.
The efficient frontier is a hyperbola where the horizontal axis represents risk (standard deviation) and the vertical axis represents expected return.

The portfolio which produces the leftmost point on this hyperbola is the global variance-minimal portfolio $p_*$ with return and variance $(\mu_*, \sigma^2_*)$.
Its return splits up the curve into two parts, $\mu<\mu_*$ and $\mu \geq \mu_*$.
The upper part of this hyperbola consists of efficient portfolios whereas the lower part consists of inefficient portfolios, because for these, a portfolio with a higher return and the same variance also exists.
Now, if we select $\epsilon \geq \mu_*$, the minimum variance portfolio for a larger-than-$\epsilon$ return has exactly return $\epsilon$.
In our benchmark test set we select $\epsilon$ as the 70 \% quantile of its asset returns.
Thus, we do not explicitly enforce $\epsilon \geq \mu_*$.
Nonetheless, we assume this to be mostly the case, because with the 70 \% quantile, we have a higher-than-average returning portfolio.
We checked this assumption by solving the ``$\mu \geq \epsilon$'' quadratic program for all instances of our benchmark set.
We observe that $93,75 \ \%$ of the variance-minimal portfolio displayed $\mu = \epsilon$.
Furthermore, we observe that this assumption tends to be more often violated on small instances.
When considering only instances that we used in the quantum benchmarks, i.e., from 5 to 30 assets, $\mu = \epsilon$ holds only with a percentage of $80 \ \%$.
To further analyze the magnitude of the incurred inaccuracy, we calculate the difference in variance between optimizing with ``$\geq$'' and ``$=$'' for the instances where $\mu = \epsilon$ did not hold.
The mean relative difference amounts to $0.025$.
To also report absolute values, the mean absolute difference in optimal variances for instances from 5 to 30 assets turns out to be $0.0015$, the maximum difference is $0.004$.
As a result of these small deviations, we argue that it is indeed reasonable to replace $\geq$ by $=$, as in
 most cases, the optimum solution to the ``$\geq$'' quadratic program is attained at $\mu = \epsilon$.
For instances where this is not the case, the incurred error in optimum variance is small.
Replacing $\geq$ by $=$ also benefits the QUBO model because no additional slack variables have to be introduced, which in turn would have to be discretized and thus would heavily increase the qubit count and further reduce the accessible instance sizes.
We acknowledge that this is a trade-off between model-size and modelling accuracy.
The same argumentation holds for a one-sided penalty construction, which introduces additional non-quadratic structure or auxiliary variables.

\section{Pseudocode Problem-specific Heuristic}\label{sec:minvola_heuristic}
The Problem-specific heuristic for the \hyperlink{eq:min_vola}{\text{MinVola}} problem formulation with $n \in \mathbb{N}$ possible assets generates up to $n$ feasible solutions.
\begin{algorithm}[H]
\caption{MinVola - Problem-specific Heuristic}
\label{alg:minvola_heuristic}
\begin{algorithmic}[1]

\State Let $\mu_i \in \mathbb{R} \gets$ expected return for asset $i \in \{1, \dots, n\}$
\State Let $u_i \in (0,1] \gets$ upper bound of asset weight for asset $i \in \{1, \dots, n\}$
\State Let $\sigma_{ij} \in \mathbb{R} \gets$ covariance of returns of assets $i,j \in \{1, \dots, n\}$
\State Let $\epsilon$ $\in \mathbb{R} \gets$ minimum required portfolio return
\State Let $\delta \in (0,1] \gets$ constant weight parameter that gets added to the samples
\State Initialize \texttt{Sols} $\gets$ empty list of solutions

\For{$i \gets 1$ \textbf{to} $n$}
 \State Initialize $x \gets [0, \dots, 0]$ empty asset weight vector of length $n$
 \State $x_i \gets \delta \quad$ add weight to asset $i$
 \While{$\sum_{k=1}^{n} x_k < 1-\frac{\delta}{2}$}
 \State $x \gets \Call{AddNewAssetWeightSteepestDesc}{x, \mu, ub, \sigma, \delta, \epsilon}$
 \EndWhile
 \State $x \gets \Call{NormalizeAssetWeights}{x}$
 \State Add $x$ to \texttt{Sols}
\EndFor
\State \Return \texttt{Sols}

\vspace{1em}
\Function{AddNewAssetWeightSteepestDesc}{$x, \mu, ub, \sigma, \delta, \epsilon$}
 \State Let $\text{big\_M} \gets 1000$
 \State Let $\text{small\_M} \gets -1000$
 \State Initialize $Variances \gets [\ ] \quad$ variances
 \State Initialize $Rets \gets [\ ] \quad $ returns
 \For{$j \gets 1$ \textbf{to} $n$}
 \State Let $x' \gets$ copy of $x$
 \State $x'_j \gets x'_j + \delta$
 \If{$x'_j \leq ub_j$}
 \State $r \gets \frac{\sum_{k=1}^n x'_k \mu_k}{\sum_{k=1}^n x'_k} \quad$ calculate normalized return
 \State Append $r$ to $Rets$
 \If{$r \geq$ $\epsilon$}
 \State v $\gets \sum_{k=1}^n \sum_{l=1}^n x'_k \sigma_{kl} x'_l \quad$ calculate return variance
 \State Append v to $Variances$
 \Else
 \State Append $\text{big\_M}$ to $Variances$
 \EndIf
 \Else
 \State Append $\text{small\_M}$ to $Rets$
 \State Append $\text{big\_M}$ to $Variances$
 \EndIf
 \EndFor
 \If{$\min($$Variances) = \text{big\_M}$}
 \State $j^* \gets \arg\max Rets$
 \Else
 \State $j^* \gets \arg\min$ $Variances$
 \EndIf
 \State $x_{j^*} \gets x_{j^*} + \delta$
 \State \Return $x$
\EndFunction

\vspace{1em}
\Function{NormalizeAssetWeights}{$x$}
\State $\text{sum\_of\_weights} \gets \sum_{i=1}^{n} x_i$
\For{$i \gets 1$ \textbf{to} $n$}
\State $x_i \gets x_i\cdot \text{sum\_of\_weights}$
\EndFor
\State \Return $x$
\EndFunction

\end{algorithmic}
\end{algorithm}

\section{Selection of Penalty Factors}\label{sec:penalty_factors}

\begin{figure}
 \centering
 \includegraphics[width=\linewidth]{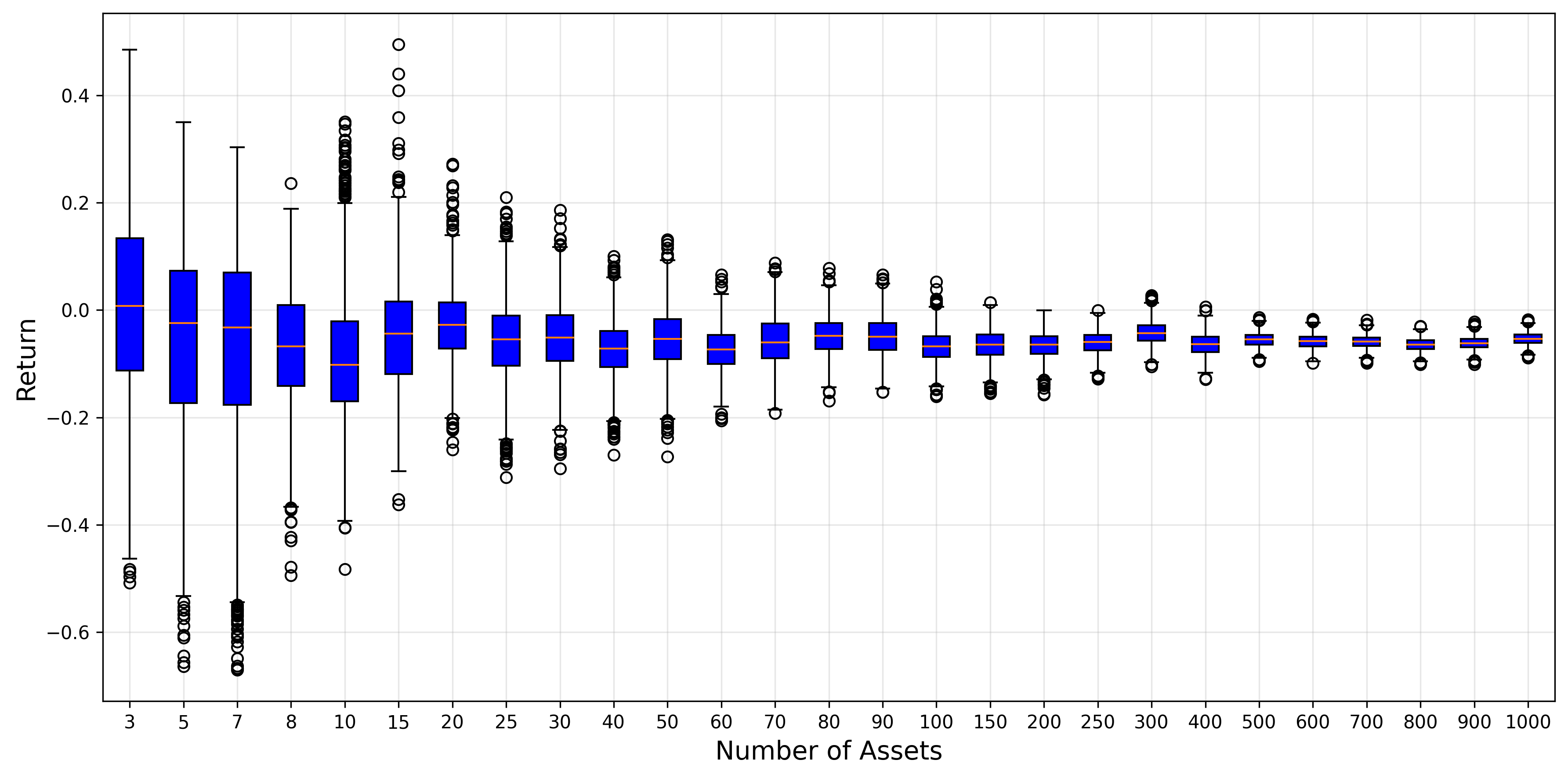}
 \caption{Distribution of returns of random portfolios for different instance sizes.
 For each instance size, we sampled $100$ random portfolios for each of the $10$ problem instances in our test data set.}
 \label{fig:rets_of_random_portfolios}
\end{figure}

\begin{figure}
 \centering
 \includegraphics[width=\linewidth]{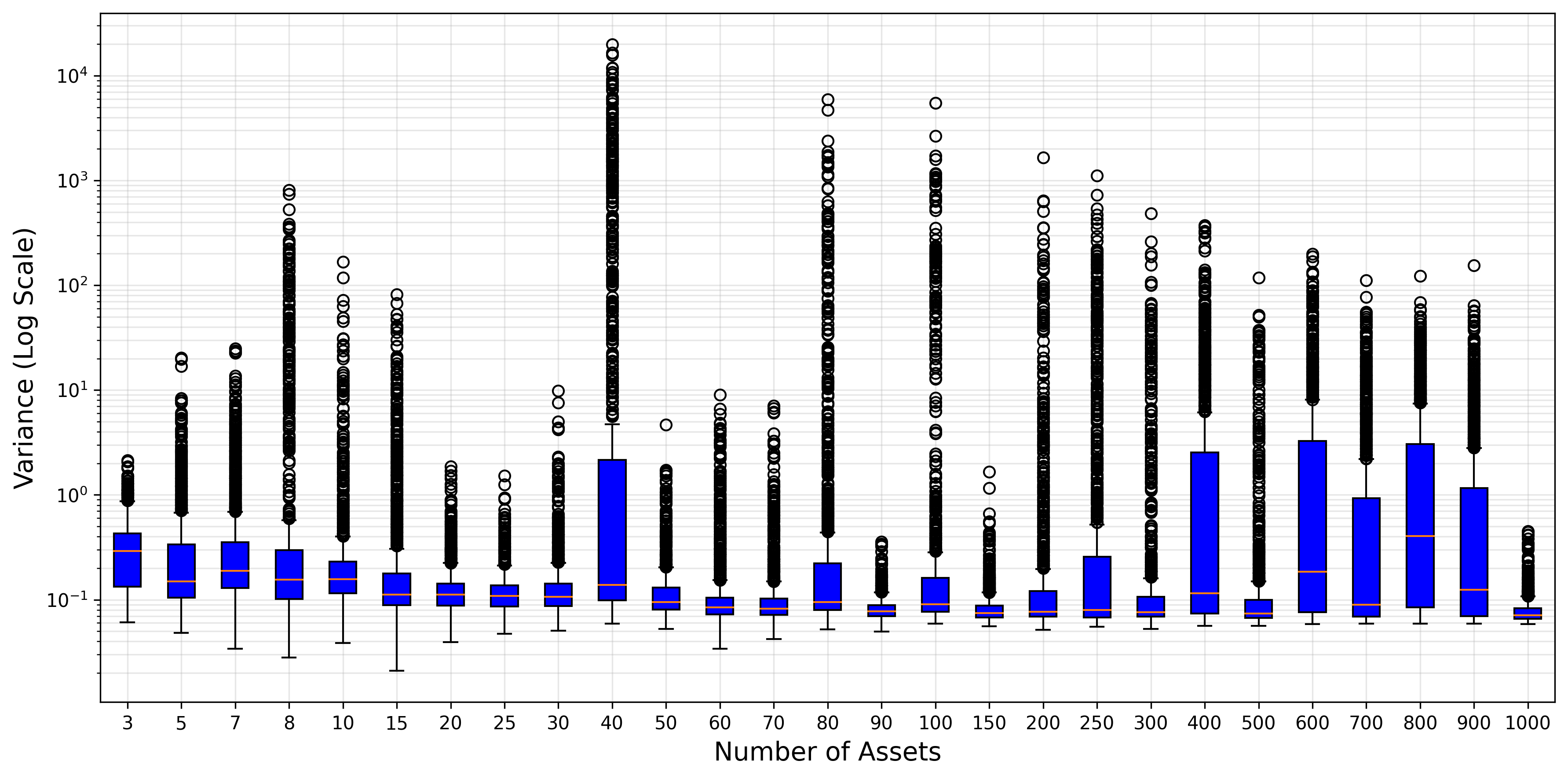}
 \caption{Distribution of variances of random portfolios for different instance sizes.
 For each instance size, we sampled $100$ random portfolios for each of the $10$ problem instances in our test data set.}
 \label{fig:volas_of_random_portfolios}
\end{figure}

In the QUBO model~\eqref{eq:qubo_obj}, we set the penalty factors $\phi=\psi=1000$ for violations of the return constraint~\eqref{eq:minreturn_constraint} and the normalization constraint~\eqref{eq:normalization_constraint}, respectively.
Thus, an absolute constraint violation of $0.01$ leads to a penalization of $1000 \cdot 0.01^2 = 0.1$.
The goal is to make constraint violations unfavorable by increasing the objective value.
Thus, we require that a constraint violation of $0.01$ is small compared to the typical right-hand side of~\eqref{eq:minreturn_constraint} and~\eqref{eq:normalization_constraint}
and, at the same time, that a penalization of $0.1$ is comparable to a typical objective value of~\hyperlink{eq:min_vola}{\text{MinVola}}.
Clearly, $0.01$ is small compared to the right-hand-side $1$ of the normalization constraint~\eqref{eq:normalization_constraint}.
To analyze the domain of the right-hand-side of the return constraint~\eqref{eq:minreturn_constraint}, i.e.\ the 70\%-quantile of the return $\mu(\omega)$, and the domain of the objective, i.e. the variance $\sigma^2(\omega)$, we calculate the returns and variances for 100 random portfolios for each of the $10$ instances for each problem size (asset number).
Thus, we analyze 28,000 random portfolios in total.
The results are visualized in the boxplots in Fig.~\ref{fig:rets_of_random_portfolios} and Fig.~\ref{fig:volas_of_random_portfolios}.

The boxplots visualize the statistical distribution of numerical data points by showing the median (red line), the 25th and 75th percentiles (blue box boundaries), the range of non-outlier values within $1.5$ times the interquartile range (whiskers), and individual outlier points outside the whiskers.
From this data, we estimate the typical magnitudes of portfolio returns and variances.
The 75th percentiles (upper boundaries) of the portfolio returns are mostly negative and take values in the range from -0.1 to 0.1.
We see that a return constraint violation of $0.01$ is indeed small compared to a typical absolute value of $\epsilon$, especially for instances with a small number of assets.
These violations happen with decreasing frequency with increasing problem size.
In particular, for $n\leq30$ assets, which is the maximum number of assets that we are able to execute on a quantum computers, these frequently appearing high violations are penalized heavily, which is desired.

\begin{figure}
 \centering
 \includegraphics[width=\linewidth]{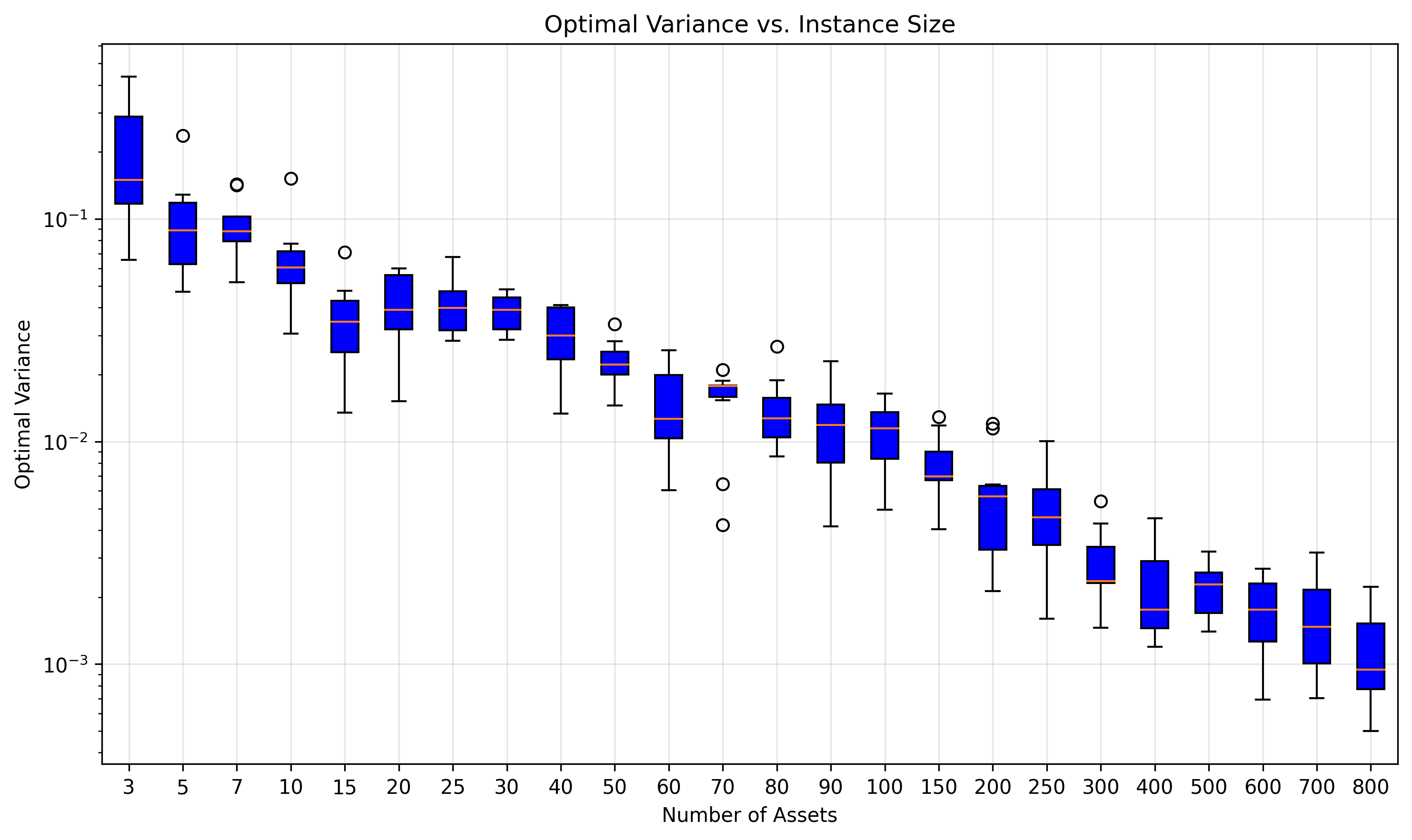}
 \caption{Distribution of optimal variances of the 10 discretized instances of each problem size in the benchmark data set.}
 \label{fig:optimal_volas}
\end{figure}

For the variances, the number of outliers is larger than for the returns.
Some outliers have values above $10,000$ (e.g. for $40$ assets).
This is due to the high variance of the return variances of the individual assets which can cause a huge return variance in a random portfolio.
Nonetheless, all medians have values below $0.35$.
Moreover, the optimal solutions to the instances show decreasing variances with increasing problem sizes, all below $0.5$.
This effect is visualized in Fig.~\ref{fig:optimal_volas}.
Thus, a penalty of $0.1$ for a return-constraint violation of $0.01$, is in a comparable range to the objective function as desired.

\section{Benchmark Study Results: Simulated QAOA}\label{sec:simqaoa_results}
\begin{figure}[p]
 \centering
 \begin{subfigure}[b]{0.9\textwidth}
 \includegraphics[width=\linewidth]{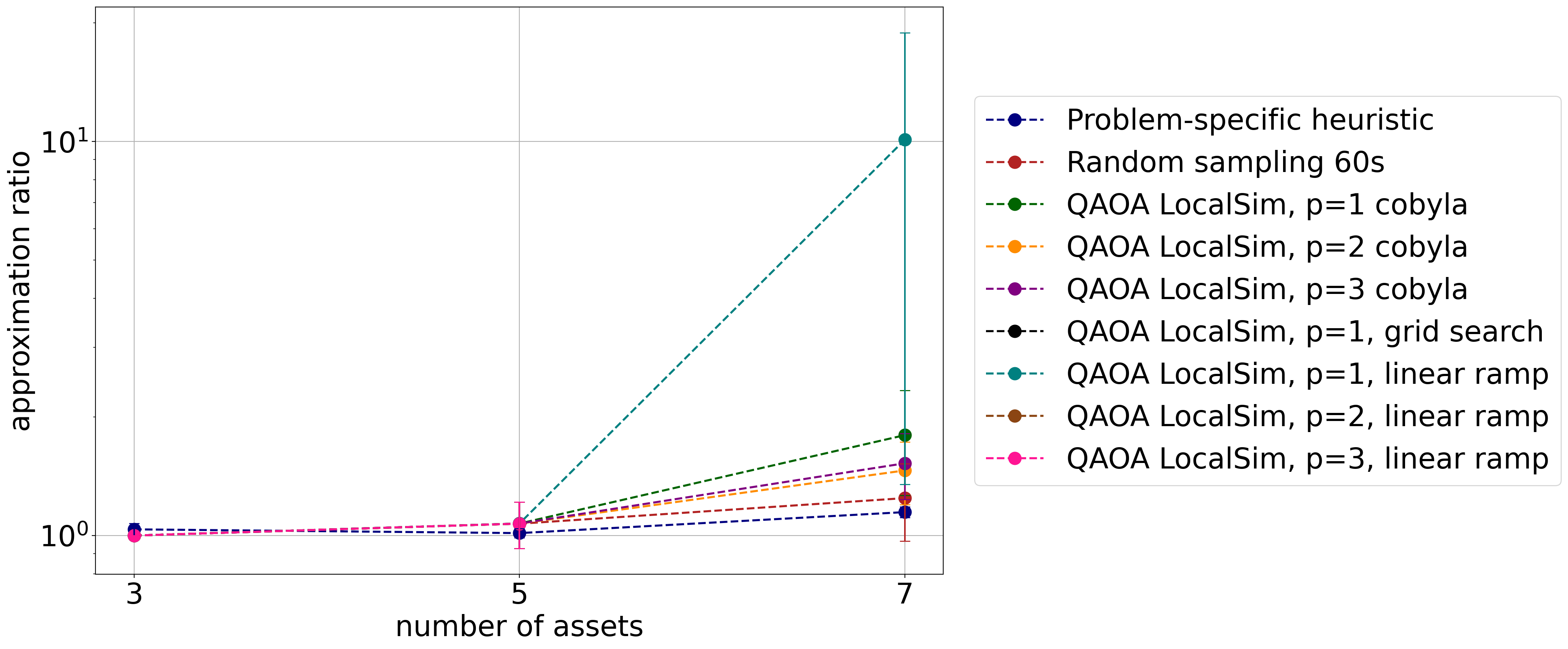}
 \caption{Simulated QAOA approximation ratio}
 \label{fig:sim_qaoa_approximation_ratio}
 \end{subfigure}
 \begin{subfigure}[b]{0.9\textwidth}
 \includegraphics[width=\linewidth]{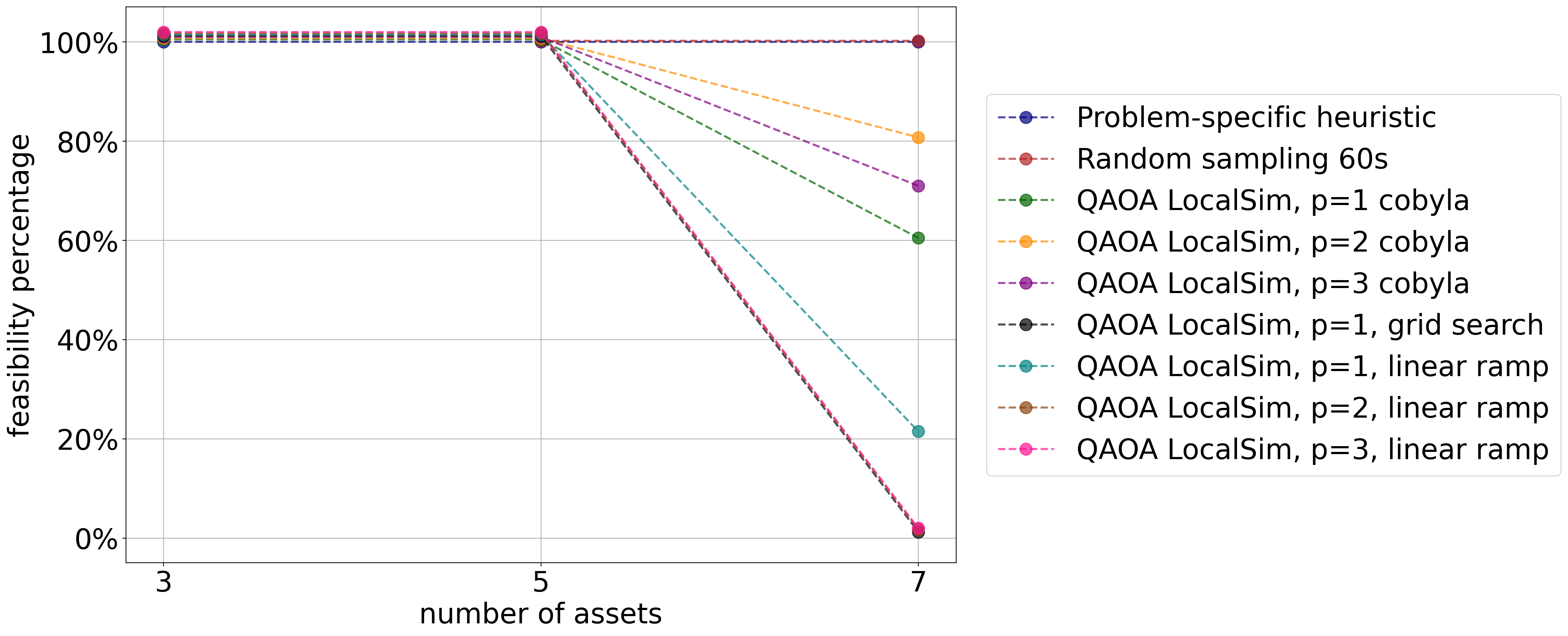}
 \caption{Simulated QAOA feasibility percentage}
 \label{fig:sim_qaoa_feasibility_percentage}
 \end{subfigure}
 \begin{subfigure}[b]{0.9\textwidth}
 \includegraphics[width=\linewidth]{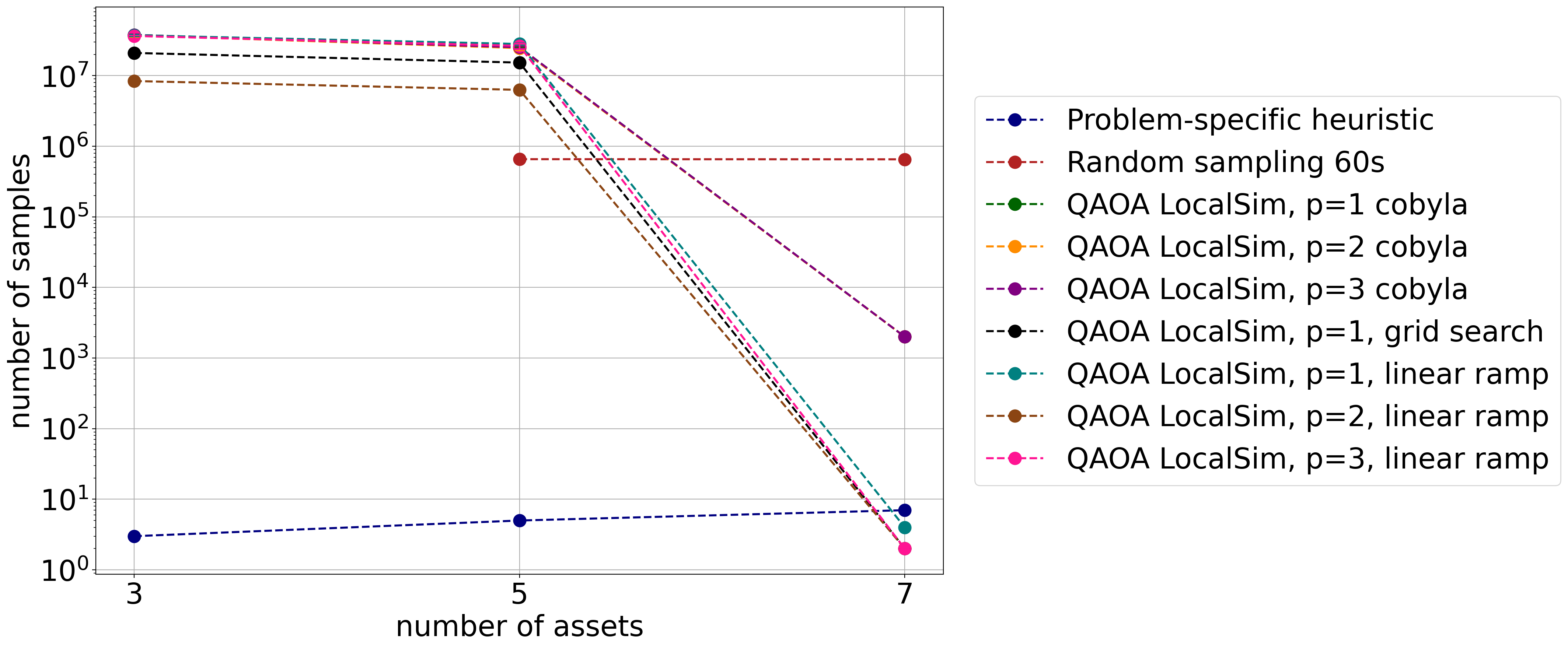}
 \caption{Simulated QAOA number of samples}
 \label{fig:sim_qaoa_number_of_shots}
 \end{subfigure}
 \caption{
 Approximation ratio, feasibility percentage and number of samples for the simulated QAOA.
 We compare against the problem-specific heuristic and random sampling as baselines.
 Data points are averages over $10$ instances and error bars show the empirical standard deviation.
 Missing data points in the approximation ratio are due to the absence of feasible solutions.
 \label{fig:sim_qaoa_results}}
\end{figure}

As mentioned in Section~\ref{sec:methods}, a classically simulated QAOA, e.g. with AerSimulator~\cite{noauthor_aersimulator_nodate}, can only be executed for up to around 30 variables.
As a result, the largest problem size that can be run with our configuration on the simulator is $7$ assets, requiring 28 variables.
For $3,5$ and $7$ assets we show the results in Fig.~\ref{fig:sim_qaoa_results}.
Therein, we compare different configurations of the locally simulated QAOA against random sampling and the problem-specific heuristic.
We differentiate between the number of layers $p=1,2,3$ and the way of optimizing the QAOA parameters.
First, we use the COBYLA optimizer to optimize the parameters classically.
Second, we perform a grid search and third, we initialize the parameters with the linear ramp formula.
For further details on the grid search and the linear ramp, we refer to the QAOA section~\ref{sec:qaoa}.

\section{Benchmark Study Results: Distribution of Approximation Ratios}\label{sec:approximationratio_boxplots}

\begin{figure}[p]
 \begin{subfigure}[b]{0.85\textwidth}
 \centering
 \includegraphics[width=\linewidth]{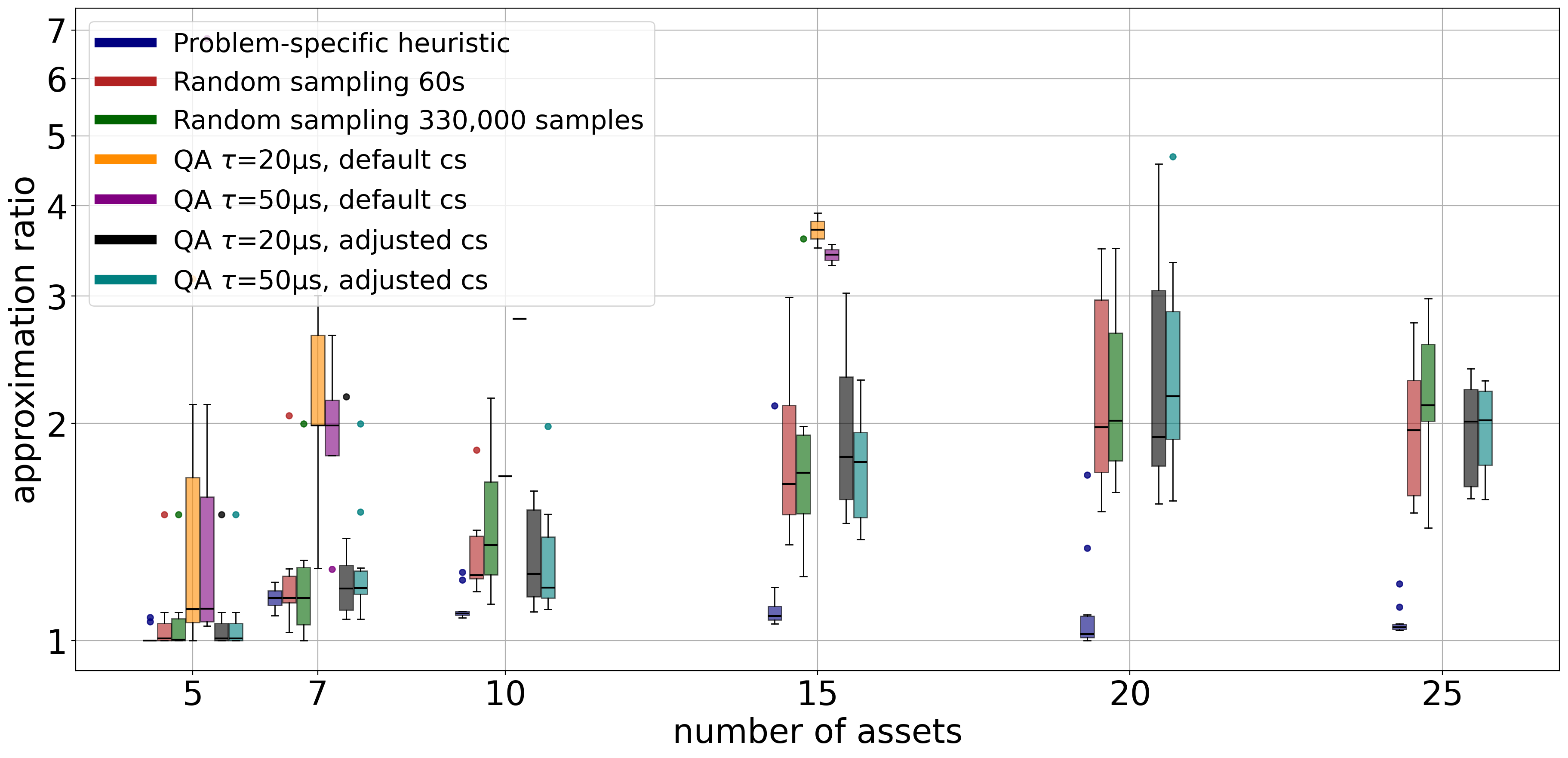}
 \caption{Boxplots of the quantum annealing approximation ratios of solving the 10 instances of each problem size in the benchmark data set.}
 \label{fig:annealing_approximation_ratio_boxplot}
 \end{subfigure}

 \begin{subfigure}[b]{0.85\textwidth}
 \centering
 \includegraphics[width=\linewidth]{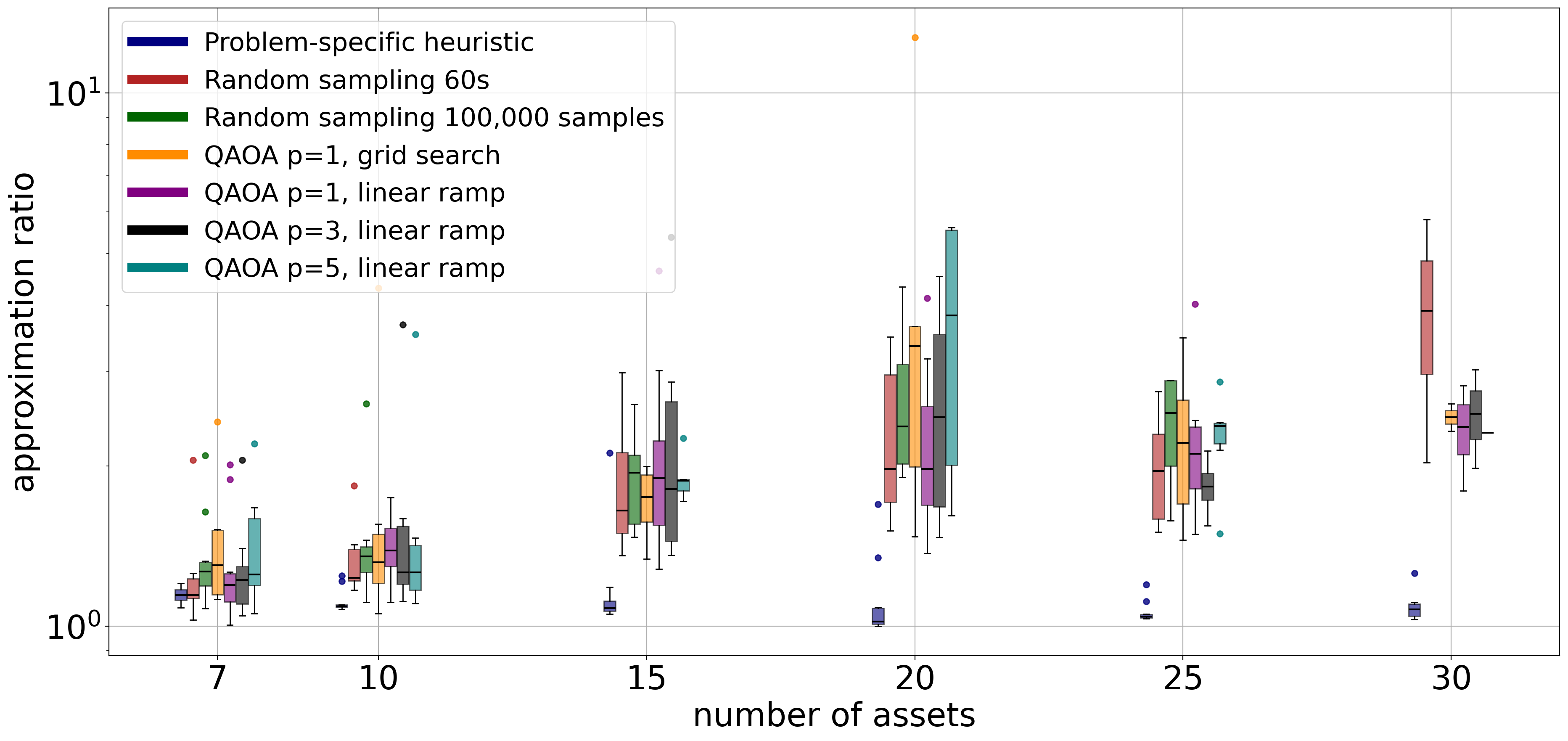}
 \caption{Boxplots of the QAOA approximation ratios of solving the 10 instances of each problem size in the benchmark data set.}
 \label{fig:qaoa_approximation_ratio_boxplot}
 \end{subfigure}

 \begin{subfigure}[b]{0.85\textwidth}
 \centering
 \includegraphics[width=\linewidth]{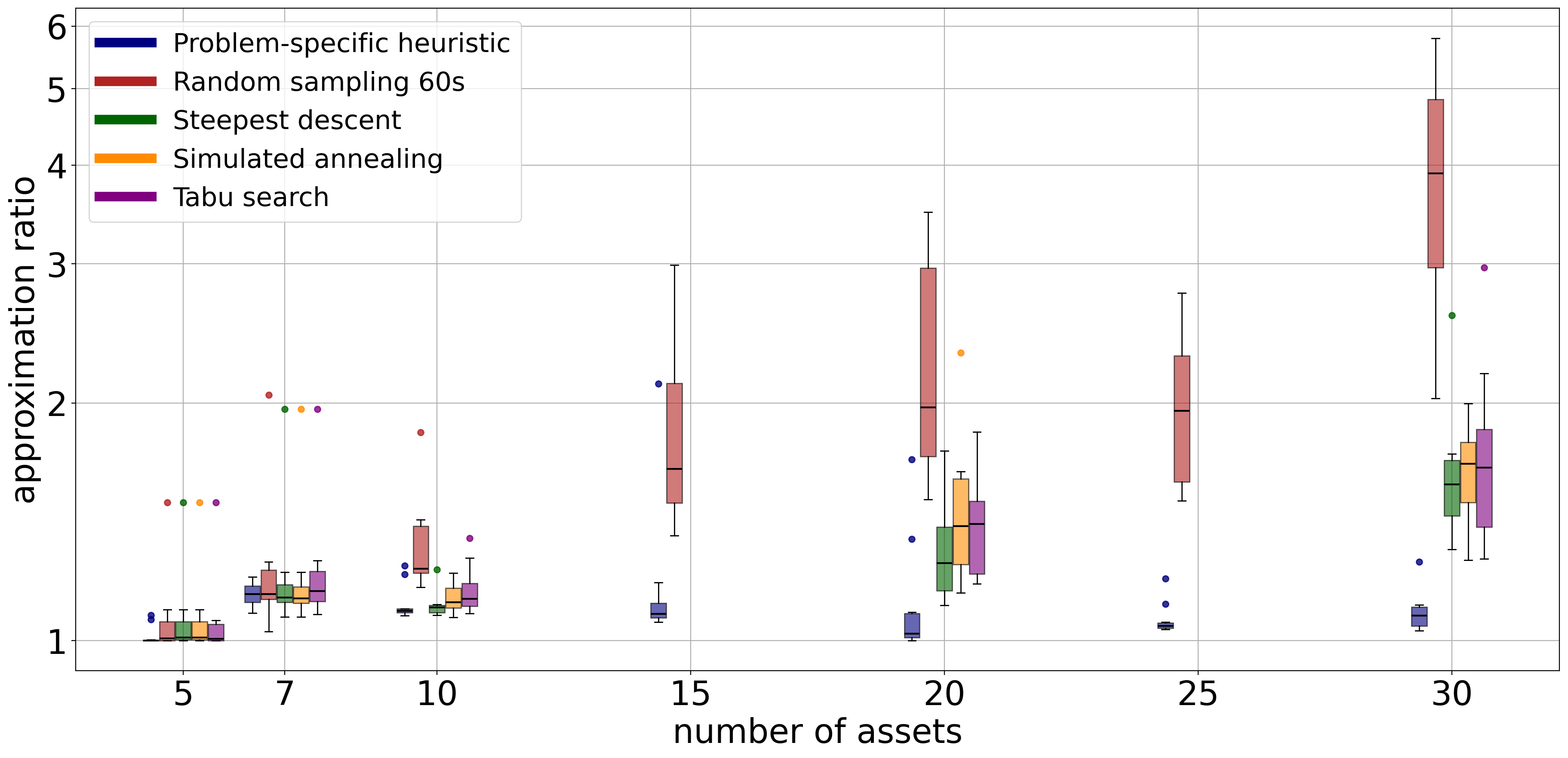}
 \caption{Boxplots of the heuristics approximation ratios of solving the 10 instances of each problem size in the benchmark data set.}
 \label{fig:heuristics_approximation_ratio_boxplot}
 \end{subfigure}
 \caption{Boxplots of the approximation ratios of quantum annealing, QAOA, and the heuristics.
 They visualize the median (black lines inside the boxes), the 25th and 75th percentiles (box boundaries), the range of non-outlier values within 1.5 times the inter-quartile range (whiskers), and individual outlier points outside the whiskers.
 }
\label{fig:boxplots_approximationratios}
\end{figure}

In the result section~\ref{sec:benchmark_study_results}, we show the mean and empirical standard deviation of the achieved approximation ratios.
In this appendix, we additionally visualize the distributions by boxplots in Fig.~\ref{fig:boxplots_approximationratios}.

\section{Benchmark Study Results: Approximation Ratios With Respect to Continuous Problem}\label{sec:approximationratio_continuous}
\begin{figure}[p]
 \begin{subfigure}[b]{0.9\textwidth}
 \centering
 \includegraphics[width=\linewidth]{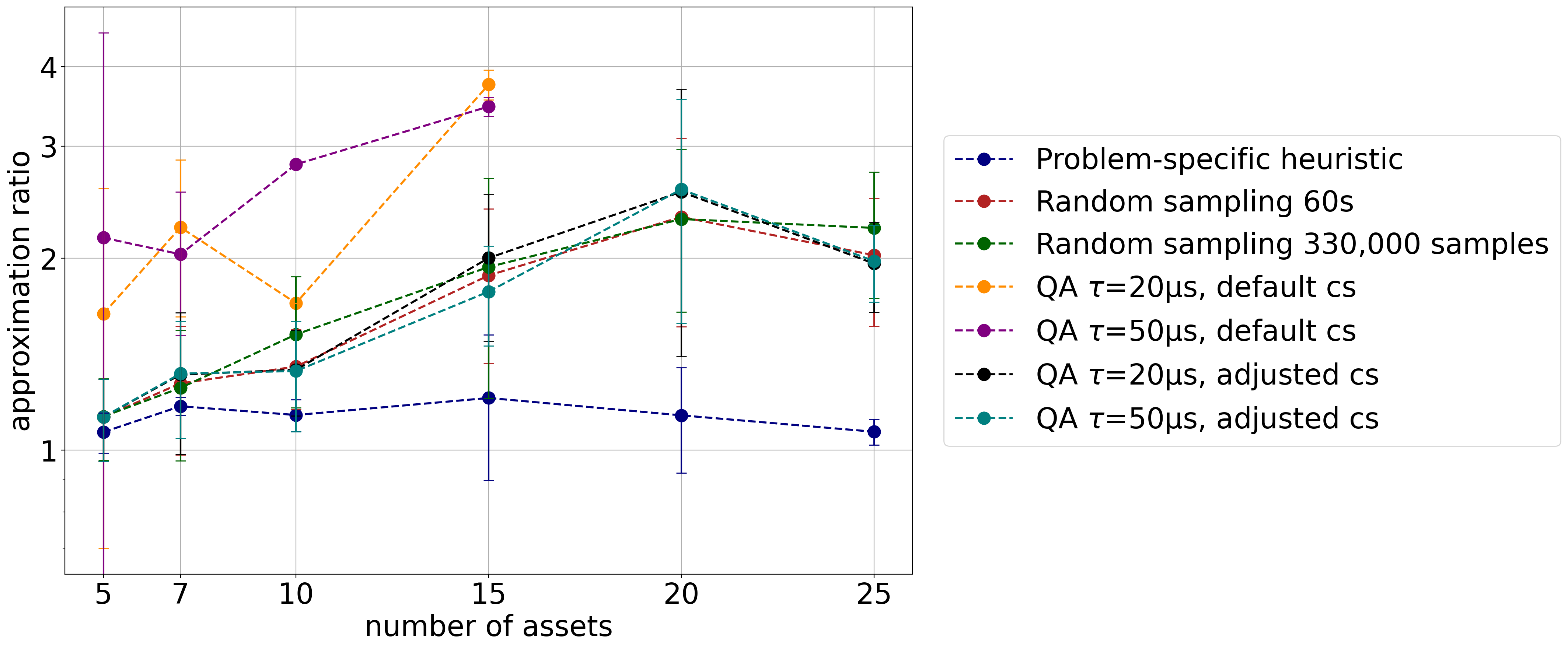}
 \caption{Quantum annealing approximation ratios w.r.t.\ the continuous \hyperlink{eq:min_vola}{\text{MinVola}} problem.} \label{fig:annealing_approximation_ratio_continuous}
 \end{subfigure}

 \begin{subfigure}[b]{0.9\textwidth}
 \centering
 \includegraphics[width=\linewidth]{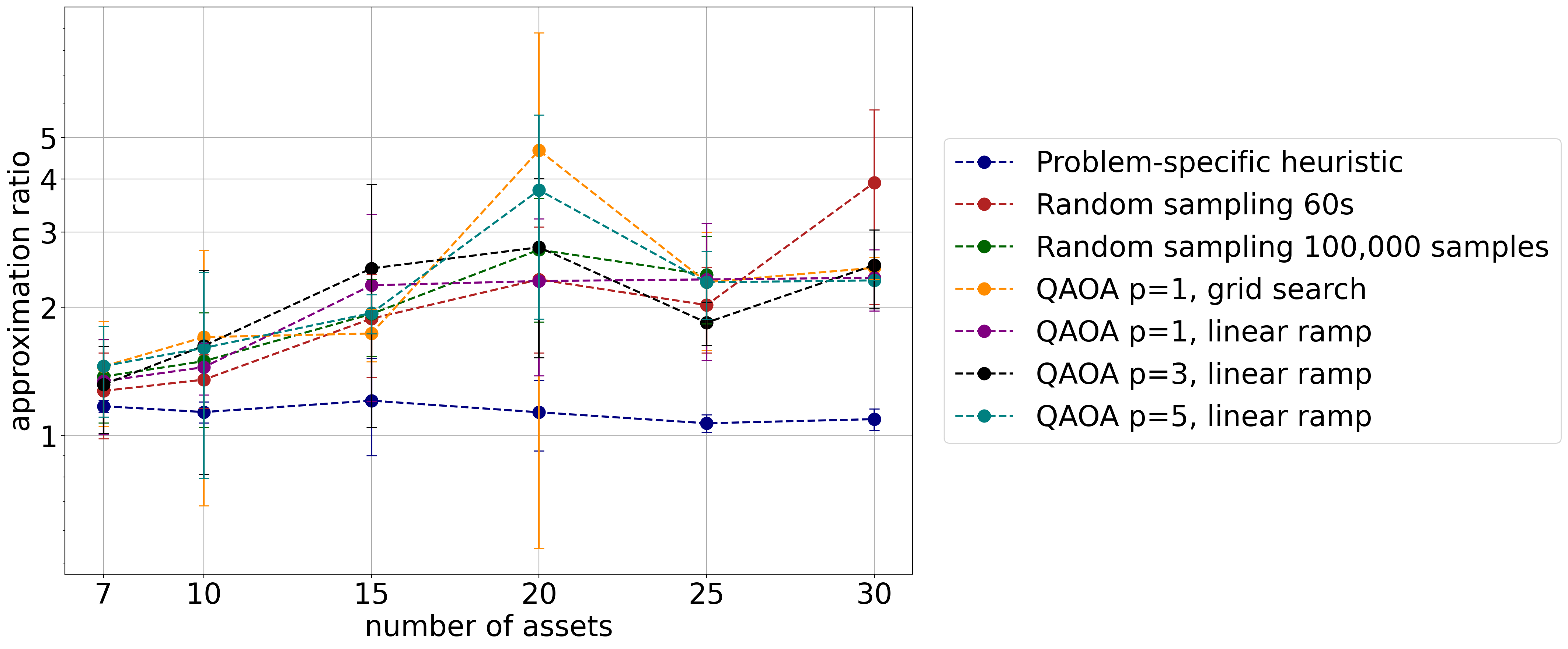}
 \caption{QAOA approximation ratios w.r.t.\ the continuous \hyperlink{eq:min_vola}{\text{MinVola}} problem.} \label{fig:qaoa_approximation_ratio_continuous}
 \end{subfigure}

 \begin{subfigure}[b]{0.9\textwidth}
 \centering
 \includegraphics[width=\linewidth]{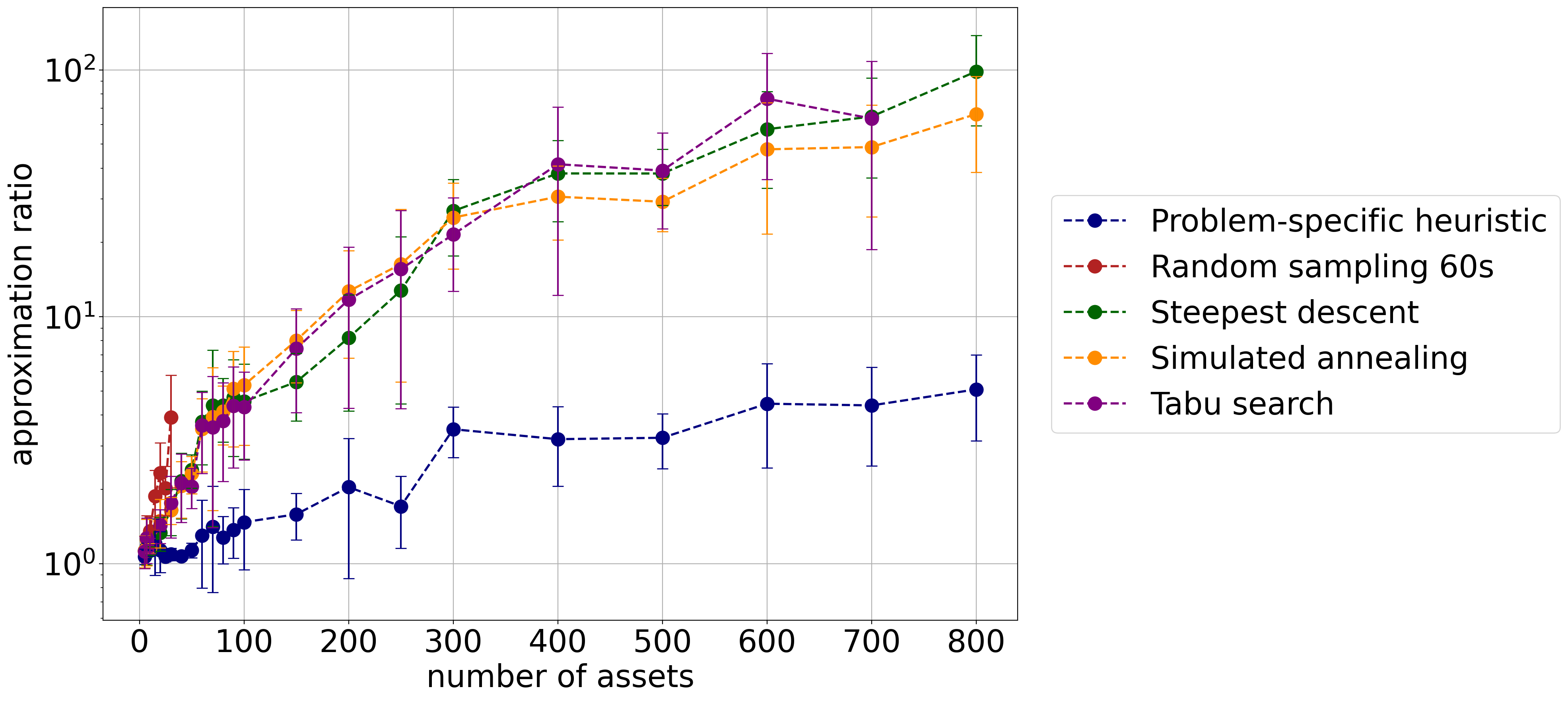}
 \caption{Heuristics approximation ratios w.r.t.\ the continuous \hyperlink{eq:min_vola}{\text{MinVola}} problem.}
 \label{fig:heuristics_approximation_ratio_continuous}
 \end{subfigure}
 \caption{The approximation ratios of quantum annealing, QAOA, and the heuristics, where $f_{opt}$ is selected as the optimal solution of the continuous version of the Minvola problem.}
 \label{fig:approximationratios_continuous}
\end{figure}

In section \ref{sec:benchmark_study_results}, we report the approximation ratios based on the objective value of the optimal solution to the discretized formulation of the \hyperlink{eq:min_vola}{\text{MinVola}} problem.
In Fig.~\ref{fig:approximationratios_continuous}, we additionally report the approximation ratios with respect to the optimal solution to the continuous \hyperlink{eq:min_vola}{\text{MinVola}} problem.
We observe similar results for the continuous case as for the discretized case.
This similarity is due to the small difference between the optimal solution values of the discretized and the continuous problem,
which further justifies the selection of the discretization factor of $d=3$.

\newpage

    \printbibliography

@article{montanez-barrera_towards_2025,
	title = {Towards a {Linear}-{Ramp} {QAOA} protocol: {Evidence} of a scaling advantage in solving some combinatorial optimization problems},
	volume = {11},
	issn = {2056-6387},
	shorttitle = {Towards a {Linear}-{Ramp} {QAOA} protocol},
	url = {http://arxiv.org/abs/2405.09169},
	doi = {10.1038/s41534-025-01082-1},
	number = {1},
	urldate = {2026-10-01},
	journal = {npj Quantum Information},
	author = {Montanez-Barrera, J. A. and Michielsen, Kristel},
	month = aug,
	year = {2025},
	note = {arXiv:2405.09169 [quant-ph]},
	pages = {131},
}

@misc{sakuler_real_2023,
	title = {A real world test of {Portfolio} {Optimization} with {Quantum} {Annealing}},
	url = {http://arxiv.org/abs/2303.12601},
	language = {en},
	urldate = {2024-06-11},
	publisher = {arXiv},
	author = {Sakuler, Wolfgang and Oberreuter, Johannes M. and Aiolfi, Riccardo and Asproni, Luca and Roman, Branislav and Schiefer, Jürgen},
	month = mar,
	year = {2023},
	note = {arXiv:2303.12601 [quant-ph]},
}

@article{ozaeta_expectation_2022,
	title = {Expectation values from the single-layer quantum approximate optimization algorithm on {Ising} problems},
	volume = {7},
	issn = {2058-9565},
	url = {https://iopscience.iop.org/article/10.1088/2058-9565/ac9013},
	language = {en},
	number = {4},
	urldate = {2024-06-11},
	journal = {Quantum Science and Technology},
	author = {Ozaeta, Asier and Van Dam, Wim and McMahon, Peter L},
	month = oct,
	year = {2022},
	pages = {045036},
}

@article{mcgeoch_milestones_2023,
	title = {Milestones on the {Quantum} {Utility} {Highway}: {Quantum} {Annealing} {Case} {Study}},
	volume = {5},
	doi = {10.1145/3625307},
	number = {1},
	journal = {ACM Transactions on Quantum Computing},
	author = {McGeoch, Catherine C. and Farré, Pau},
	month = dec,
	year = {2023},
	note = {Place: New York, NY, USA
Publisher: Association for Computing Machinery},
}

@article{koch_progress_2022,
	title = {Progress in mathematical programming solvers from 2001 to 2020},
	volume = {10},
	issn = {2192-4406},
	doi = {https://doi.org/10.1016/j.ejco.2022.100031},
	journal = {EURO Journal on Computational Optimization},
	author = {Koch, Thorsten and Berthold, Timo and Pedersen, Jaap and Vanaret, Charlie},
	year = {2022},
	pages = {100031},
}

@article{albash_demonstration_2018,
	title = {Demonstration of a {Scaling} {Advantage} for a {Quantum} {Annealer} over {Simulated} {Annealing}},
	volume = {8},
	doi = {10.1103/PhysRevX.8.031016},
	number = {3},
	journal = {Phys. Rev. X},
	author = {Albash, Tameem and Lidar, Daniel A.},
	month = jul,
	year = {2018},
	note = {Publisher: American Physical Society},
	pages = {031016},
}

@article{gleixner_miplib_2021,
	title = {{MIPLIB} 2017: {Data}-{Driven} {Compilation} of the 6th {Mixed}-{Integer} {Programming} {Library}},
	volume = {13},
	doi = {10.1007/s12532-020-00194-3},
	number = {3},
	journal = {Mathematical Programming Computation},
	author = {Gleixner, Ambros and Hendel, Gregor and Gamrath, Gerald and Achterberg, Tobias and Bastubbe, Michael and Berthold, Timo and Christophel, Philipp M. and Jarck, Kati and Koch, Thorsten and Linderoth, Jeff and Lübbecke, Marco and Mittelmann, Hans D. and Ozyurt, Derya and Ralphs, Ted K. and Salvagnin, Domenico and Shinano, Yuji},
	month = sep,
	year = {2021},
	pages = {443--490},
}

@article{harrigan_quantum_2021,
	title = {Quantum approximate optimization of non-planar graph problems on a planar superconducting processor},
	volume = {17},
	issn = {1745-2481},
	doi = {10.1038/s41567-020-01105-y},
	number = {3},
	journal = {Nature Physics},
	author = {Harrigan, Matthew P. and Sung, Kevin J. and Neeley, Matthew and Satzinger, Kevin J. and Arute, Frank and Arya, Kunal and Atalaya, Juan and Bardin, Joseph C. and Barends, Rami and Boixo, Sergio and Broughton, Michael and Buckley, Bob B. and Buell, David A. and Burkett, Brian and Bushnell, Nicholas and Chen, Yu and Chen, Zijun and Chiaro, Ben and Collins, Roberto and Courtney, William and Demura, Sean and Dunsworth, Andrew and Eppens, Daniel and Fowler, Austin and Foxen, Brooks and Gidney, Craig and Giustina, Marissa and Graff, Rob and Habegger, Steve and Ho, Alan and Hong, Sabrina and Huang, Trent and Ioffe, L. B. and Isakov, Sergei V. and Jeffrey, Evan and Jiang, Zhang and Jones, Cody and Kafri, Dvir and Kechedzhi, Kostyantyn and Kelly, Julian and Kim, Seon and Klimov, Paul V. and Korotkov, Alexander N. and Kostritsa, Fedor and Landhuis, David and Laptev, Pavel and Lindmark, Mike and Leib, Martin and Martin, Orion and Martinis, John M. and McClean, Jarrod R. and McEwen, Matt and Megrant, Anthony and Mi, Xiao and Mohseni, Masoud and Mruczkiewicz, Wojciech and Mutus, Josh and Naaman, Ofer and Neill, Charles and Neukart, Florian and Niu, Murphy Yuezhen and O’Brien, Thomas E. and O’Gorman, Bryan and Ostby, Eric and Petukhov, Andre and Putterman, Harald and Quintana, Chris and Roushan, Pedram and Rubin, Nicholas C. and Sank, Daniel and Skolik, Andrea and Smelyanskiy, Vadim and Strain, Doug and Streif, Michael and Szalay, Marco and Vainsencher, Amit and White, Theodore and Yao, Z. Jamie and Yeh, Ping and Zalcman, Adam and Zhou, Leo and Neven, Hartmut and Bacon, Dave and Lucero, Erik and Farhi, Edward and Babbush, Ryan},
	month = feb,
	year = {2021},
	note = {Publisher: Springer Science and Business Media LLC},
	pages = {332--336},
}

@article{acuaviva_benchmarking_2026,
	title = {Benchmarking quantum computers: towards a standard performance evaluation approach},
	volume = {11},
	issn = {2058-9565},
	shorttitle = {Benchmarking quantum computers},
	url = {https://doi.org/10.1088/2058-9565/ae4269},
	doi = {10.1088/2058-9565/ae4269},
	language = {en},
	number = {2},
	urldate = {2026-10-01},
	journal = {Quantum Science and Technology},
	publisher = {IOP Publishing},
	author = {Acuaviva, Arturo and Aguirre, David and Peña, Rubén and Sanz, Mikel},
	month = feb,
	year = {2026},
	pages = {025004},
}

@article{venturelli_reverse_2019,
	title = {Reverse {Quantum} {Annealing} {Approach} to {Portfolio} {Optimization} {Problems}},
	volume = {1},
	issn = {2524-4906, 2524-4914},
	url = {http://arxiv.org/abs/1810.08584},
	doi = {10.1007/s42484-019-00001-w},
	number = {1-2},
	urldate = {2026-10-01},
	journal = {Quantum Machine Intelligence},
	author = {Venturelli, Davide and Kondratyev, Alexei},
	month = may,
	year = {2019},
	note = {arXiv:1810.08584 [quant-ph]},
	pages = {17--30},
}

@misc{palmer_quantum_2021,
	title = {Quantum {Portfolio} {Optimization} with {Investment} {Bands} and {Target} {Volatility}},
	url = {http://arxiv.org/abs/2106.06735},
	urldate = {2024-12-30},
	publisher = {arXiv},
	author = {Palmer, Samuel and Sahin, Serkan and Hernandez, Rodrigo and Mugel, Samuel and Orus, Roman},
	month = aug,
	year = {2021},
	note = {arXiv:2106.06735 [q-fin]},
}

@article{brandhofer_benchmarking_2022,
	title = {Benchmarking the performance of portfolio optimization with {QAOA}},
	volume = {22},
	issn = {1573-1332},
	url = {http://arxiv.org/abs/2207.10555},
	doi = {10.1007/s11128-022-03766-5},
	number = {1},
	urldate = {2024-12-30},
	journal = {Quantum Information Processing},
	author = {Brandhofer, Sebastian and Braun, Daniel and Dehn, Vanessa and Hellstern, Gerhard and Hüls, Matthias and Ji, Yanjun and Polian, Ilia and Bhatia, Amandeep Singh and Wellens, Thomas},
	month = dec,
	year = {2022},
	note = {arXiv:2207.10555 [quant-ph]},
	pages = {25},
}

@article{acharya_decomposition_2025,
	title = {Decomposition {Pipeline} for {Large}-{Scale} {Portfolio} {Optimization} with {Applications} to {Near}-{Term} {Quantum} {Computing}},
	volume = {7},
	issn = {2643-1564},
	url = {http://arxiv.org/abs/2409.10301},
	doi = {10.1103/PhysRevResearch.7.023142},
	number = {2},
	urldate = {2026-10-01},
	journal = {Physical Review Research},
	author = {Acharya, Atithi and Yalovetzky, Romina and Minssen, Pierre and Chakrabarti, Shouvanik and Shaydulin, Ruslan and Raymond, Rudy and Sun, Yue and Herman, Dylan and Andrist, Ruben S. and Salton, Grant and Schuetz, Martin J. A. and Katzgraber, Helmut G. and Pistoia, Marco},
	month = may,
	year = {2025},
	note = {arXiv:2409.10301 [math.OC]},
	pages = {023142},
}

@misc{rubio-garcia_portfolio_2022,
	title = {Portfolio optimization with discrete simulated annealing},
	url = {http://arxiv.org/abs/2210.00807},
	urldate = {2024-12-30},
	publisher = {arXiv},
	author = {Rubio-García, Álvaro and García-Ripoll, Juan José and Porras, Diego},
	month = oct,
	year = {2022},
	note = {arXiv:2210.00807 [cond-mat]},
}

@misc{buhler_efficient_2023,
	title = {Efficient {Solution} of {Portfolio} {Optimization} {Problems} via {Dimension} {Reduction} and {Sparsification}},
	url = {http://arxiv.org/abs/2306.12639},
	urldate = {2025-01-07},
	publisher = {arXiv},
	author = {Buhler, Cassidy K. and Benson, Hande Y.},
	month = jun,
	year = {2023},
	note = {arXiv:2306.12639 [q-fin]},
}

@article{loke_portfolio_2023,
	title = {Portfolio {Optimization} {Problem}: {A} {Taxonomic} {Review} of {Solution} {Methodologies}},
	volume = {11},
	copyright = {https://creativecommons.org/licenses/by/4.0/legalcode},
	issn = {2169-3536},
	shorttitle = {Portfolio {Optimization} {Problem}},
	url = {https://ieeexplore.ieee.org/document/10087257/},
	language = {en},
	urldate = {2025-01-07},
	journal = {IEEE Access},
	author = {Loke, Zi Xuan and Goh, Say Leng and Kendall, Graham and Abdullah, Salwani and Sabar, Nasser R.},
	year = {2023},
	pages = {33100--33120},
}

@article{cesarone_efficient_nodate,
	title = {Efficient {Algorithms} for mean-variance portfolio optimization with {Hard} {Real} -{World} {Constraints}},
	language = {en},
	author = {Cesarone, Francesco and Scozzari, Andrea and Tardella, Fabio},
    journal={Giornale dell'Istituto Italiano degli Attuari},
    year = {2009},
    pages = {37--56},
}

@misc{noauthor_leader_nodate,
	title = {The {Leader} in {Decision} {Intelligence} {Technology} - {Gurobi} {Optimization}},
	url = {https://www.gurobi.com/},
	urldate = {2025-03-04},
}

@article{markowitz_portfolio_1952,
	title = {Portfolio {Selection}},
	volume = {7},
	copyright = {© 1952 the American Finance Association},
	issn = {1540-6261},
	url = {https://onlinelibrary.wiley.com/doi/abs/10.1111/j.1540-6261.1952.tb01525.x},
	language = {en},
	number = {1},
	urldate = {2025-03-05},
	journal = {The Journal of Finance},
	author = {Markowitz, Harry},
	year = {1952},
	pages = {77--91},
}

@misc{ammann_realistic_2023,
	title = {Realistic {Runtime} {Analysis} for {Quantum} {Simplex} {Computation}},
	url = {http://arxiv.org/abs/2311.09995},
	urldate = {2025-05-05},
	publisher = {arXiv},
	author = {Ammann, Sabrina and Hess, Maximilian and Ramacciotti, Debora and Fekete, Sándor P. and Goedicke, Paulina L. A. and Gross, David and Lefterovici, Andreea and Osborne, Tobias J. and Perk, Michael and Rotundo, Antonio and Skelton, S. E. and Stiller, Sebastian and Wolff, Timo de},
	month = nov,
	year = {2023},
	note = {arXiv:2311.09995 [quant-ph]},
}

@article{bochkarev_quantum_2026,
	title = {Quantum computing for discrete optimization: {A} highlight of three technologies},
	volume = {329},
	issn = {0377-2217},
	url = {https://www.sciencedirect.com/science/article/pii/S0377221725005880},
	doi = {https://doi.org/10.1016/j.ejor.2025.07.063},
	number = {3},
	journal = {European Journal of Operational Research},
	author = {Bochkarev, Alexey and Heese, Raoul and Jäger, Sven and Schiewe, Philine and Schöbel, Anita},
	year = {2026},
	pages = {747--766},
}

@article{koch_quantum_2026,
	title = {Quantum {Optimization} {Benchmarking} {Library} - {The} {Intractable} {Decathlon}},
	volume = {6},
	issn = {2662-8457},
	url = {http://arxiv.org/abs/2504.03832},
	doi = {10.1038/s43588-026-00991-1},
	number = {6},
	urldate = {2026-10-01},
	journal = {Nature Computational Science},
	author = {Koch, Thorsten and Neira, David E. Bernal and Chen, Ying and Cortiana, Giorgio and Egger, Daniel J. and Heese, Raoul and Hegade, Narendra N. and Cadavid, Alejandro Gomez and Huang, Rhea and Itoko, Toshinari and Kleinert, Thomas and Xavier, Pedro Maciel and Mohseni, Naeimeh and Montanez-Barrera, Jhon A. and Nakano, Koji and Nannicini, Giacomo and O'Meara, Corey and Pauckert, Justin and Proissl, Manuel and Ramesh, Anurag and Schicker, Maximilian and Shimada, Noriaki and Takeori, Mitsuharu and Valls, Victor and Bulck, David Van and Woerner, Stefan and Zoufal, Christa},
	month = jun,
	year = {2026},
	note = {arXiv:2504.03832 [quant-ph]},
	pages = {653--671},
}

@article{Kirkpatrick1983,
  author = {Kirkpatrick, S. and Gelatt, C. D. and Vecchi, M. P.},
  copyright = {Copyright © 1983 American Association for the Advancement of Science},
  description = {JSTOR: Science, New Series, Vol. 220, No. 4598 (May 13, 1983), pp. 671-680},
  doi = {10.1126/science.220.4598.671},
  issn = {00368075},
  journal = {Science},
  jstor_articletype = {primary_article},
  jstor_formatteddate = {May 13, 1983},
  number = 4598,
  pages = {671--680},
  publisher = {American Association for the Advancement of Science},
  series = {New Series},
  title = {Optimization by Simulated Annealing},
  url = {http://www.jstor.org/stable/1690046},
  volume = 220,
  year = 1983
}

@article{abughanem_ibm_2025,
	title = {{IBM} {Quantum} {Computers}: {Evolution}, {Performance}, and {Future} {Directions}},
	volume = {81},
	issn = {1573-0484},
	shorttitle = {{IBM} {Quantum} {Computers}},
	url = {http://arxiv.org/abs/2410.00916},
	doi = {10.1007/s11227-025-07047-7},
	number = {5},
	urldate = {2026-10-01},
	journal = {The Journal of Supercomputing},
	author = {AbuGhanem, M.},
	month = apr,
	year = {2025},
	note = {arXiv:2410.00916 [quant-ph]},
	pages = {687},
}

@article{merton_efficient_frontier_1972,
 ISSN = {00221090, 17566916},
 URL = {http://www.jstor.org/stable/2329621},
 author = {Robert C. Merton},
 journal = {The Journal of Financial and Quantitative Analysis},
 number = {4},
 pages = {1851--1872},
 publisher = {Cambridge University Press},
 title = {An Analytic Derivation of the Efficient Portfolio Frontier},
 urldate = {2026-07-03},
 volume = {7},
 year = {1972}
}

@misc{schlütter2025hotstartingquantumportfoliooptimization,
      title={Hot-Starting Quantum Portfolio Optimization}, 
      author={Sebastian Schlütter and Tomislav Maras and Alexander Dotterweich and Nico Piatkowski},
      year={2025},
      eprint={2510.11153},
      archivePrefix={arXiv},
      primaryClass={quant-ph},
      url={https://arxiv.org/abs/2510.11153}, 
}

@article{Scursulim_2026,
   title={Multiclass portfolio optimization via variational quantum Eigensolver with Dicke state ansatz},
   volume={16},
   ISSN={2045-2322},
   url={http://dx.doi.org/10.1038/s41598-026-36333-4},
   DOI={10.1038/s41598-026-36333-4},
   number={1},
   journal={Scientific Reports},
   publisher={Springer Science and Business Media LLC},
   author={Scursulim, J. V. S. and Langeloh, Gabriel M. and Beltran, Victor L. and Brito, Samuraí},
   year={2026},
   month=Feb }

@article{Egger_2021,
   title={Warm-starting quantum optimization},
   volume={5},
   ISSN={2521-327X},
   url={http://dx.doi.org/10.22331/q-2021-06-17-479},
   DOI={10.22331/q-2021-06-17-479},
   journal={Quantum},
   publisher={Verein zur Forderung des Open Access Publizierens in den Quantenwissenschaften},
   author={Egger, Daniel J. and Mareček, Jakub and Woerner, Stefan},
   year={2021},
   month=June, pages={479} }

@article{Khezri_2022,
   title={Customized Quantum Annealing Schedules},
   volume={17},
   ISSN={2331-7019},
   url={http://dx.doi.org/10.1103/PhysRevApplied.17.044005},
   DOI={10.1103/physrevapplied.17.044005},
   number={4},
   journal={Physical Review Applied},
   publisher={American Physical Society (APS)},
   author={Khezri, Mostafa and Dai, Xi and Yang, Rui and Albash, Tameem and Lupascu, Adrian and Lidar, Daniel A.},
   year={2022},
   month=Apr }

@misc{noauthor_nasdaq_nodate,
	title = {{Stock Screener NASDAQ}},
	url = {https://www.nasdaq.com/market-activity/stocks/screener},
	language = {en},
	urldate = {2026-05-26}
}

@misc{noauthor_yfinance_2025,
	title = {yfinance: Python Package to download market data from Yahoo! Finance API},
	url = {https://pypi.org/project/yfinance/0.2.66/},
	urldate = {2026-10-01},
	month = sep,
	year = {2025},
}

@misc{noauthor_pyscipopt_2025,
	title = {{PySCIPOpt} ({Python} {Package} for {SCIP})},
	url = {https://pypi.org/project/PySCIPOpt/5.6.0/},
	language = {de},
	urldate = {2026-10-01},
	journal = {PyPI},
	month = aug,
	year = {2025},
}

@misc{noauthor_dwave_system_2025,
	title = {dwave-system ({Python} {Package} for {Dwave})},
	url = {https://pypi.org/project/dwave-system/1.33.0/},
	language = {de},
	urldate = {2026-10-01},
	journal = {PyPI},
	month = sep,
	year = {2025},
}

@misc{noauthor_qiskit_aer_2025,
	title = {qiskit-aer ({Python} {Package} for quantum circuit simulation)},
	url = {https://pypi.org/project/qiskit-aer/0.17.2/},
	language = {de},
	urldate = {2026-10-01},
	journal = {PyPI},
	month = sep,
	year = {2025},
}

@misc{noauthor_scipy_2025,
	title = {scipy (open-source software for mathematics, science, and engineering)\#},
	url = {https://pypi.org/project/scipy/1.16.2/},
	language = {de},
	urldate = {2026-10-01},
	journal = {PyPI},
	month = sep,
	year = {2025},
}

@misc{noauthor_minimizemethodcobyla_nodate,
	title = {minimize(method=’{COBYLA}’) — {SciPy} v1.15.3 {Manual}},
	url = {https://docs.scipy.org/doc/scipy/reference/optimize.minimize-cobyla.html},
	urldate = {2025-05-30},
}

@misc{noauthor_aersimulator_nodate,
	title = {{AerSimulator} - {Qiskit} {Aer} 0.16.1},
	url = {https://qiskit.github.io/qiskit-aer/stubs/qiskit_aer.AerSimulator.html},
	urldate = {2025-05-30},
}

@misc{noauthor_dwave_greedy_2022,
	title = {dwave-greedy (Python Package for Dwave Greedy)},
	url = {https://pypi.org/project/dwave-greedy/0.3.0/},
	language = {de},
	urldate = {2026-10-01},
	journal = {PyPI},
	month = nov,
	year = {2022},
}

@misc{noauthor_dwave_tabu_2022,
	title = {dwave-tabu (Python Package for Dwave Tabu)},
	url = {https://pypi.org/project/dwave-tabu/0.5.0/},
	language = {de},
	urldate = {2026-10-01},
	journal = {PyPI},
	month = nov,
	year = {2022},
}

@misc{noauthor_dwave_samplers_2025,
	title = {dwave-samplers ({Python} {Package} for {Dwave} samplers)},
	url = {https://pypi.org/project/dwave-samplers/1.7.0/},
	language = {de},
	urldate = {2026-10-01},
	journal = {PyPI},
	month = oct,
	year = {2025},
}

@book{boyd_convex_2004,
	address = {Cambridge, UK ; New York},
	title = {Convex optimization},
	isbn = {978-0-521-83378-3},
	language = {en},
	publisher = {Cambridge University Press},
	author = {Boyd, Stephen P. and Vandenberghe, Lieven},
	year = {2004},
}

@book{messiah_quantum_1961,
	title = {Quantum {Mechanics}},
	isbn = {978-0-7204-0045-8},
	language = {en},
	publisher = {Elsevier},
	author = {Messiah, Albert},
	year = {1961},
	note = {Google-Books-ID: VR93vUk8d\_8C},
}

@article{hauke_perspectives_2020,
	title = {Perspectives of quantum annealing: {Methods} and implementations},
	volume = {83},
	issn = {0034-4885, 1361-6633},
	shorttitle = {Perspectives of quantum annealing},
	url = {http://arxiv.org/abs/1903.06559},
	doi = {10.1088/1361-6633/ab85b8},
	number = {5},
	urldate = {2025-05-30},
	journal = {Reports on Progress in Physics},
	author = {Hauke, Philipp and Katzgraber, Helmut G. and Lechner, Wolfgang and Nishimori, Hidetoshi and Oliver, William D.},
	month = may,
	year = {2020},
	note = {arXiv:1903.06559 [quant-ph]},
	pages = {054401},
}

@misc{noauthor_dwavesystemsdwave-greedy_2024,
	title = {dwavesystems/dwave-greedy},
	copyright = {Apache-2.0},
	url = {https://github.com/dwavesystems/dwave-greedy},
	urldate = {2025-05-30},
	publisher = {D-Wave Quantum Inc.},
	month = sep,
	year = {2024},
	note = {original-date: 2019-08-02T10:10:41Z},
}

@misc{noauthor_dwavesystemsdwave-tabu_2024,
	title = {dwavesystems/dwave-tabu},
	copyright = {Apache-2.0},
	url = {https://github.com/dwavesystems/dwave-tabu},
	urldate = {2025-05-30},
	publisher = {D-Wave Quantum Inc.},
	month = sep,
	year = {2024},
	note = {original-date: 2018-08-28T02:14:30Z},
}

@misc{noauthor_dwave-samplers_nodate,
	title = {dwave-samplers — {Python} documentation},
	url = {https://docs.dwavequantum.com/en/latest/ocean/api_ref_samplers/index.html},
	urldate = {2025-05-30},
}

@article{chang_heuristics_2000,
	title = {Heuristics for cardinality constrained portfolio optimisation},
	volume = {27},
	copyright = {https://www.elsevier.com/tdm/userlicense/1.0/},
	issn = {03050548},
	url = {https://linkinghub.elsevier.com/retrieve/pii/S030505489900074X},
	language = {en},
	number = {13},
	urldate = {2025-06-05},
	journal = {Computers \& Operations Research},
	author = {Chang, T.-J. and Meade, N. and Beasley, J.E. and Sharaiha, Y.M.},
	month = nov,
	year = {2000},
	pages = {1271--1302},
}

@misc{soloviev_scaling_2025,
	title = {Scaling {Portfolio} {Diversification} with {Quantum} {Circuit} {Cutting} {Techniques}},
	url = {http://arxiv.org/abs/2506.08947},
	urldate = {2025-06-13},
	publisher = {arXiv},
	author = {Soloviev, Vicente P. and Romero, Antonio Márquez and Kirsopp, Josh and Krompiec, Michal},
	month = jun,
	year = {2025},
	note = {arXiv:2506.08947 [quant-ph]},
}

@misc{moeini_continuous_2014,
	title = {A {Continuous} {Optimization} {Approach} for the {Financial} {Portfolio} {Selection} under {Discrete} {Asset} {Choice} {Constraints}},
	url = {http://arxiv.org/abs/1404.3286},
	urldate = {2025-06-13},
	publisher = {arXiv},
	author = {Moeini, Mahdi},
	month = apr,
	year = {2014},
	note = {arXiv:1404.3286 [cs]},
}

@misc{moka_scalable_2025,
	title = {A {Scalable} {Gradient}-{Based} {Optimization} {Framework} for {Sparse} {Minimum}-{Variance} {Portfolio} {Selection}},
	url = {http://arxiv.org/abs/2505.10099},
	urldate = {2025-06-13},
	publisher = {arXiv},
	author = {Moka, Sarat and Quiroz, Matias and Asimit, Vali and Muller, Samuel},
	month = may,
	year = {2025},
	note = {arXiv:2505.10099 [stat]},
}

@article{cesarone_portfolio_2015,
	title = {Portfolio selection problems in practice: a comparison between linear and quadratic optimization models},
	volume = {12},
	issn = {1619-697X, 1619-6988},
	shorttitle = {Portfolio selection problems in practice},
	url = {http://arxiv.org/abs/1105.3594},
	doi = {10.1007/s10287-014-0210-1},
	number = {3},
	urldate = {2025-06-13},
	journal = {Computational Management Science},
	author = {Cesarone, Francesco and Scozzari, Andrea and Tardella, Fabio},
	month = jul,
	year = {2015},
	note = {arXiv:1105.3594 [q-fin]},
	pages = {345--370},
}

@misc{noauthor_yahoo_nodate,
	title = {Yahoo {Finance} - {Stock} {Market} {Live}, {Quotes}, {Business} \& {Finance} {News}},
	url = {https://finance.yahoo.com/},
	language = {en-US},
	urldate = {2025-06-13},
	journal = {Yahoo Finance},
}

@misc{bolusani_scip_nodate,
	title = {The {SCIP} {Optimization} {Suite} 9.0},
	url = {https://arxiv.org/abs/2402.17702v2},
    urldate = {2026-05-26},
    year = {2024},
	publisher = {arXiv},
    note = {arXiv:2402.17702 [math.OC]},
	language = {en},
	author = {Bolusani, Suresh and Besancon, Mathieu and Bestuzheva, Ksenia and Chmiela, Antonia and Dionısio, Joao and Donkiewicz, Tim and van Doornmalen, Jasper and Eiﬂer, Leon and Ghannam, Mohammed and Gleixner, Ambros and Graczyk, Christoph and Halbig, Katrin and Hedtke, Ivo and Hoen, Alexander and Hojny, Christopher},
    
}

@misc{noauthor_samplers_nodate,
	title = {Samplers — dwave-system 0.9.12 documentation},
	url = {https://dwave-systemdocs.readthedocs.io/en/latest/reference/samplers.html#dwavecliquesampler},
	urldate = {2025-06-24},
}

@inproceedings{phillipson_portfolio_2021,
	address = {Cham},
	title = {Portfolio {Optimisation} {Using} the {D}-{Wave} {Quantum} {Annealer}},
	isbn = {978-3-030-77980-1},
	doi = {10.1007/978-3-030-77980-1_4},
	language = {en},
	booktitle = {Computational {Science} – {ICCS} 2021},
	publisher = {Springer International Publishing},
	author = {Phillipson, Frank and Bhatia, Harshil Singh},
	editor = {Paszynski, Maciej and Kranzlmüller, Dieter and Krzhizhanovskaya, Valeria V. and Dongarra, Jack J. and Sloot, Peter M. A.},
	year = {2021},
	pages = {45--59},
}

@article{alessandroni_alleviating_2025,
	title = {Alleviating the quantum {Big}-{M} problem},
	volume = {11},
	issn = {2056-6387},
	url = {https://www.nature.com/articles/s41534-025-01067-0},
	language = {en},
	number = {1},
	urldate = {2025-07-30},
	journal = {npj Quantum Information},
	author = {Alessandroni, Edoardo and Ramos-Calderer, Sergi and Roth, Ingo and Traversi, Emiliano and Aolita, Leandro},
	month = jul,
	year = {2025},
	pages = {125},
}

@article{buonaiuto_best_2023,
	title = {Best practices for portfolio optimization by quantum computing, experimented on real quantum devices},
	volume = {13},
	issn = {2045-2322},
	url = {https://www.nature.com/articles/s41598-023-45392-w},
	language = {en},
	number = {1},
	urldate = {2025-08-04},
	journal = {Scientific Reports},
	author = {Buonaiuto, Giuseppe and Gargiulo, Francesco and De Pietro, Giuseppe and Esposito, Massimo and Pota, Marco},
	month = nov,
	year = {2023},
	pages = {19434},
}

@misc{noauthor_ibm_2022,
	title = {{IBM} {ILOG} {CPLEX} {Optimization} {Studio}},
	copyright = {© Copyright IBM Corporation 2022},
	url = {https://www.ibm.com/docs/en/icos/22.1.1?topic=cplex-meet},
	language = {en-US},
	urldate = {2025-08-21},
	month = dec,
	year = {2022},
}

@misc{hodson_portfolio_2019,
	title = {Portfolio rebalancing experiments using the {Quantum} {Alternating} {Operator} {Ansatz}},
	url = {http://arxiv.org/abs/1911.05296},
	urldate = {2025-08-21},
	publisher = {arXiv},
	author = {Hodson, Mark and Ruck, Brendan and Ong, Hugh and Garvin, David and Dulman, Stefan},
	month = nov,
	year = {2019},
	note = {arXiv:1911.05296 [quant-ph]},
}

@misc{noauthor_stopferericportfolio_opt_benchmark_nodate,
	title = {stopfereric/portfolio\_opt\_benchmark},
	url = {https://github.com/stopfereric/portfolio_opt_benchmark},
	urldate = {2025-09-22},
}

@book{korte_combinatorial_2018,
	address = {Berlin, Heidelberg},
	series = {Algorithms and {Combinatorics}},
	title = {Combinatorial {Optimization}: {Theory} and {Algorithms}},
	volume = {21},
	copyright = {https://www.springer.com/tdm},
	isbn = {978-3-662-56038-9 978-3-662-56039-6},
	shorttitle = {Combinatorial {Optimization}},
	url = {https://link.springer.com/10.1007/978-3-662-56039-6},
	language = {en},
	urldate = {2025-09-22},
	publisher = {Springer},
	author = {Korte, Bernhard and Vygen, Jens},
	year = {2018},
}

@article{abbas_challenges_2024,
	title = {Challenges and opportunities in quantum optimization},
	volume = {6},
	copyright = {2024 IBM, under exclusive licence to Springer Nature Limited},
	issn = {2522-5820},
	url = {https://www.nature.com/articles/s42254-024-00770-9},
	language = {en},
	number = {12},
	urldate = {2025-09-22},
	journal = {Nature Reviews Physics},
	author = {Abbas, Amira and Ambainis, Andris and Augustino, Brandon and Bärtschi, Andreas and Buhrman, Harry and Coffrin, Carleton and Cortiana, Giorgio and Dunjko, Vedran and Egger, Daniel J. and Elmegreen, Bruce G. and Franco, Nicola and Fratini, Filippo and Fuller, Bryce and Gacon, Julien and Gonciulea, Constantin and Gribling, Sander and Gupta, Swati and Hadfield, Stuart and Heese, Raoul and Kircher, Gerhard and Kleinert, Thomas and Koch, Thorsten and Korpas, Georgios and Lenk, Steve and Marecek, Jakub and Markov, Vanio and Mazzola, Guglielmo and Mensa, Stefano and Mohseni, Naeimeh and Nannicini, Giacomo and O’Meara, Corey and Tapia, Elena Peña and Pokutta, Sebastian and Proissl, Manuel and Rebentrost, Patrick and Sahin, Emre and Symons, Benjamin C. B. and Tornow, Sabine and Valls, Víctor and Woerner, Stefan and Wolf-Bauwens, Mira L. and Yard, Jon and Yarkoni, Sheir and Zechiel, Dirk and Zhuk, Sergiy and Zoufal, Christa},
	month = dec,
	year = {2024},
	note = {Publisher: Nature Publishing Group},
	pages = {718--735},
}

@misc{liu_hybrid_2022,
	title = {Hybrid {Gate}-{Based} and {Annealing} {Quantum} {Computing} for {Large}-{Size} {Ising} {Problems}},
	url = {http://arxiv.org/abs/2208.03283},
	urldate = {2025-09-22},
	publisher = {arXiv},
	author = {Liu, Chen-Yu and Goan, Hsi-Sheng},
	month = aug,
	year = {2022},
	note = {arXiv:2208.03283 [quant-ph]},
}

@misc{tang_comparative_2024,
	title = {Comparative analysis of diverse methodologies for portfolio optimization leveraging quantum annealing techniques},
	url = {http://arxiv.org/abs/2403.02599},
	urldate = {2025-09-22},
	publisher = {arXiv},
	author = {Tang, Zhijie and Dou, Alex Lu and Bishwas, Arit Kumar},
	month = jul,
	year = {2024},
	note = {arXiv:2403.02599 [quant-ph]},
}

@article{wright_interior-point_2004,
	title = {The interior-point revolution in optimization: {History}, recent developments, and lasting consequences},
	volume = {42},
	copyright = {https://www.ams.org/publications/copyright-and-permissions},
	issn = {0273-0979, 1088-9485},
	shorttitle = {The interior-point revolution in optimization},
	url = {https://www.ams.org/bull/2005-42-01/S0273-0979-04-01040-7/},
	language = {en},
	number = {1},
	urldate = {2025-09-22},
	journal = {Bulletin of the American Mathematical Society},
	author = {Wright, Margaret},
	month = sep,
	year = {2004},
	pages = {39--56},
}

@article{albash_adiabatic_2018,
	title = {Adiabatic quantum computation},
	volume = {90},
	doi = {10.1103/RevModPhys.90.015002},
	number = {1},
	journal = {Reviews of Modern Physics},
	author = {Albash, Tameem and Lidar, Daniel},
	year = {2018},
}

@article{choi_minor-embedding_2008,
	title = {Minor-{Embedding} in {Adiabatic} {Quantum} {Computation}: {I}. {The} {Parameter} {Setting} {Problem}},
	shorttitle = {Minor-{Embedding} in {Adiabatic} {Quantum} {Computation}},
	url = {http://arxiv.org/abs/0804.4884},
	doi = {10.48550/arXiv.0804.4884},
    journal = {Quantum Information Processing, 7},
	urldate = {2026-10-01},
	author = {Choi, Vicky},
	month = apr,
	year = {2008},
	note = {arXiv:0804.4884 [quant-ph]},
	pages = {pp193--209},
}

@article{choi_minor-embedding_2011,
	title = {Minor-embedding in adiabatic quantum computation: {II}. {Minor}-universal graph design},
	volume = {10},
	issn = {1570-0755, 1573-1332},
	shorttitle = {Minor-embedding in adiabatic quantum computation},
	url = {http://arxiv.org/abs/1001.3116},
	number = {3},
	urldate = {2025-09-22},
	journal = {Quantum Information Processing},
	author = {Choi, Vicky},
	month = jun,
	year = {2011},
	note = {arXiv:1001.3116 [quant-ph]},
	pages = {343--353},
}

@misc{noauthor_advantage_nodate,
	title = {The {Advantage}™ {Quantum} {Computer} {\textbar} {D}-{Wave}},
	url = {https://www.dwavequantum.com/solutions-and-products/systems/},
	language = {en-US},
	urldate = {2025-09-22},
}

@misc{cai_practical_2014,
	title = {A practical heuristic for finding graph minors},
	url = {http://arxiv.org/abs/1406.2741},
	urldate = {2025-09-22},
	publisher = {arXiv},
	author = {Cai, Jun and Macready, William G. and Roy, Aidan},
	month = jun,
	year = {2014},
	note = {arXiv:1406.2741 [quant-ph]},
}

@book{junger_50_2010,
	address = {Berlin, Heidelberg},
	title = {50 {Years} of {Integer} {Programming} 1958-2008: {From} the {Early} {Years} to the {State}-of-the-{Art}},
	copyright = {http://www.springer.com/tdm},
	isbn = {978-3-540-68274-5 978-3-540-68279-0},
	shorttitle = {50 {Years} of {Integer} {Programming} 1958-2008},
	url = {http://link.springer.com/10.1007/978-3-540-68279-0},
	language = {en},
	urldate = {2025-09-22},
	publisher = {Springer},
	editor = {Jünger, Michael and Liebling, Thomas M. and Naddef, Denis and Nemhauser, George L. and Pulleyblank, William R. and Reinelt, Gerhard and Rinaldi, Giovanni and Wolsey, Laurence A.},
	year = {2010},
}

@misc{farhi_quantum_2014,
	title = {A {Quantum} {Approximate} {Optimization} {Algorithm}},
	url = {http://arxiv.org/abs/1411.4028},
	urldate = {2025-09-22},
	publisher = {arXiv},
	author = {Farhi, Edward and Goldstone, Jeffrey and Gutmann, Sam},
	month = nov,
	year = {2014},
	note = {arXiv:1411.4028 [quant-ph]},
}

@misc{noauthor_generate_preset_pass_manager_nodate,
	title = {generate\_preset\_pass\_manager (v2.0)},
	url = {https://quantum.cloud.ibm.com/docs/en/api/qiskit/2.0/quantum.cloud.ibm.com/docs/en/api/qiskit/qiskit.transpiler.generate_preset_pass_manager},
	language = {en},
	urldate = {2026-09-21},
	journal = {IBM Quantum Documentation},
}
	
\end{document}